\documentclass[hyper,letterpaper]{JHEP3}
\usepackage{amsmath,amssymb,amsfonts,bm}
\usepackage{graphicx}
\usepackage{multirow}
\usepackage{verbatim}
\usepackage{appendix}

\newcommand{\bea}{\begin{eqnarray}}
\newcommand{\eea}{\end{eqnarray}}
\newcommand{\be}{\begin{equation}}
\newcommand{\ee}{\end{equation}}

\newcommand{\overcrossing}{{\,\raisebox{-.13cm}{\includegraphics[width=.5cm]{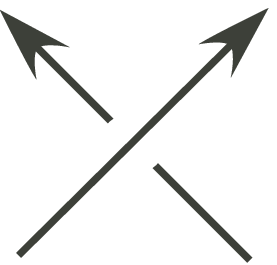}}\,}}
\newcommand{\undercrossing}{{\,\raisebox{-.13cm}{\includegraphics[width=.5cm]{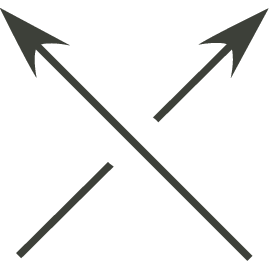}}\,}}
\newcommand{\smoothing}{{\,\raisebox{-.13cm}{\includegraphics[width=.5cm]{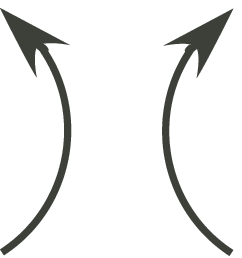}}\,}}
\newcommand{\unknot}{{\,\raisebox{-.08cm}{\includegraphics[width=.4cm]{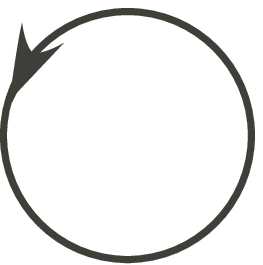}}\,}}

\newcommand{\Z}{{\mathbb Z}}
\newcommand{\R}{{\mathbb R}}
\newcommand{\C}{{\mathbb C}}

\newcommand{\Q}{{\mathbb Q}}

\newcommand{\cN}{{\mathcal{N}}}
\newcommand{\Li}{{\rm Li}}
\newcommand{\D}{{\frak D}}

\def\Tr{{\rm Tr \,}}

\def\m{\mu}

\newcommand{\cC}{{\cal C }}

\renewcommand{\hat}{\widehat}

\title{A-polynomial, B-model, and Quantization}

\author{Sergei Gukov$^{1,2}$ and Piotr Su{\l}kowski$^{1,3}$
\\ ~
\\
$^1$ California Institute of Technology, Pasadena, CA 91125, USA \\
$^2$ Max-Planck-Institut f\"ur Mathematik, Vivatsgasse 7, D-53111 Bonn, Germany \\
$^3$ Faculty of Physics, University of Warsaw, ul. Ho{\.z}a 69, 00-681 Warsaw, Poland}

\abstract{Exact solution to many problems in mathematical physics and quantum field theory
often can be expressed in terms of an algebraic curve equipped with a meromorphic differential.
Typically, the geometry of the curve can be seen most clearly in a suitable semi-classical limit,
as $\hbar \to 0$, and becomes non-commutative or ``quantum'' away from this limit.
For a classical curve defined by the zero locus of a polynomial $A(x,y)$,
we provide a construction of its non-commutative counterpart $\widehat{A} (\hat x, \hat y)$
using the technique of the topological recursion. 
This leads to a powerful and systematic algorithm for computing~$\widehat{A}$ that,
surprisingly, turns out to be much simpler than any of the existent methods.
In particular, as a bonus feature of our approach comes a curious observation that,
for all curves that come from knots or topological strings, their non-commutative counterparts
can be determined just from the first few steps of the topological recursion.
We also propose a K-theory criterion for a curve to be ``quantizable,'' and then
apply our construction to many examples that come from applications to knots, strings, instantons, and random matrices.
\\
\\
\\
\\
\\
\\
}

\preprint{CALT-68-2842}

\begin{document}


\section{Introduction}  \label{sec-intro}

In recent years, it has been realized that a solution to a variety of different
problems in theoretical and mathematical physics --- matrix models,
four-dimensional supersymmetric gauge theory, quantum invariants of knots and 3-manifolds,
and topological strings --- leads to what sometimes
is referred to as the ``quantization of an algebraic curve.''

To be more precise, the classical phase space which is quantized in this problem
is the two-dimensional complex plane parametrized by the coordinates $u$ and $v$
\be
(u,v) \; \in \; \C \times \C \,,
\ee
and equipped with the canonical holomorphic symplectic form
\be
\omega \; = \; \frac{i}{\hbar} du \wedge dv \,.
\label{sympform}
\ee
In this space, a polynomial $A(u,v)$ defines an algebraic curve
\be
\cC: \quad A(u,v) \; = \; 0 \,,
\label{class-curve}
\ee
which is automatically Lagrangian with respect to the holomorphic
symplectic form \eqref{sympform}.
A close cousin of this problem (that we consider in parallel)
is obtained by taking $A$ to be a polynomial in the $\C^*$-valued variables
\be
x = e^u \qquad , \qquad y = e^v \,.
\ee
In either case, the problem is to quantize the classical phase space
$\C \times \C$ (resp. $\C^* \times \C^*$) with the symplectic form \eqref{sympform}
and a classical ``state'' defined by the zero locus of the polynomial~$A$.

Classically, $u$ and $v$ have the Poisson bracket $\{ v,u \} = \hbar$ that follows
directly from \eqref{sympform}.
Quantization turns $u$ and $v$ into operators, $\hat{u}$ and $\hat{v}$,
which satisfy the commutation relation
\be
[\hat{v} , \hat{u}] = \hbar \,.
\label{uvcomm}
\ee
Therefore, quantization deforms the algebra of functions on the phase space into
a non-commutative algebra of operators.
In particular, it maps a polynomial function $A (u,v)$
(resp. $A(x,y)$) into an operator $\widehat{A}$:
\be
\widehat{A} \; = \; \widehat{A}_0 + \hbar \widehat{A}_1 + \hbar^2 \widehat{A}_2 + \ldots \,,
\label{Ahatpert}
\ee
where $\widehat{A}_0 \equiv A$.
Since $\hat{u}$ and $\hat{v}$ (resp. $\hat{x}$ and $\hat{y}$)
do not commute, there is no unique way to write the perturbative expansion \eqref{Ahatpert}.
After all, changing the order of operators changes the powers of $\hbar$. 
In practice, however, one often makes a choice of polarization,
{\it i.e.} a choice of what one regards as canonical coordinates and conjugate momenta.
For example, in most of the present paper we make a simple choice
consistent with~\eqref{uvcomm}:
\be
\hat{u} = u
\qquad , \qquad
\hat{v} = \hbar \partial_u \equiv \hbar \frac{\partial}{\partial u} \,,
\label{uvchoice}
\ee
where $u$ plays the role of a ``coordinate'' and $v$ is the ``momentum.''
With this or any other choice, one has a natural ordering of operators in \eqref{Ahatpert},
such that in every term momenta appear to the right of the coordinates.
This leads to a ``canonical'' form of the perturbative expansion \eqref{Ahatpert}
that we will try to follow in the present paper.

\begin{table}[h]
\centering
\begin{tabular}{|@{$\Bigm|$}c|c@{$\Bigm|$}c|c@{$\Bigm|$}|}
\hline
\textbf{Model} & \textbf{Classical curve}  & \textbf{Quantum operator}\\
\hline
\hline
Airy & $v^2 - u$ & $\hat{v}^2 - \hat{u}$ \\
\hline
tetrahedron & $1 + y + x y^f$ & $1 + q^{-1/2}\hat{y} + q^{(f+1)/2} \hat{x} \hat{y}^f$ \\
\hline
$c=1$ model & $u^2 - v^2 + 2t $ & $\hat{u}^2 - \hat{v}^2 +2t + \hbar$ \\
\hline
conifold & $ 1 + x + y + e^t xy^{-1}$ &  $1 + q^{1/2}\hat x  + q^{-1/2} \hat y + e^t \hat x  \hat{y}^{-1}$ \\
\hline
$(p,q)$ minimal & $v^p - u^q$ & $?$ \\
model & & \\
\hline
figure-$8$ & $ (1 - x^2 - 2x^4 - x^6 + x^8) y$~~ &
~$(1 - q^4 \hat{x}^4) (1 - q^2 \hat{x}^2 - (q^2 + q^6) \hat{x}^4 - q^6 \hat{x}^6 + q^8 \hat{x}^8) \hat{y}$~ \\
knot & $- x^4 - x^4 y^2$ &
$- q^{3} (1-q^6 \hat{x}^4) \hat{x}^4
- q^{5} (1 - q^2 \hat{x}^4) \hat{x}^4 \hat{y}^2$ \\
\hline
\end{tabular}
\caption{Classical $A$-polynomial and its quantization in prominent examples.\label{table} }
\end{table}

Starting with the classical curve \eqref{class-curve} defined by the zero locus of $A(u,v)$ or $A(x,y)$,
our goal will be to construct the quantum operator $\widehat{A}$,
in particular, to study the structure of its perturbative expansion \eqref{Ahatpert}.
A priori, it is not even clear if a solution to this problem exists and, if it does, whether it is unique.
We will answer these questions in affirmative and describe a systematic method
to produce ``quantum corrections'' $\widehat{A}_k$, for $k \ge 1$, solely from
the data of $A(u,v)$ (resp. $A(x,y)$)
by drawing important lessons from applications where this problem naturally appears:

\begin{itemize}

\item[\bf 1.] {\bf SUSY gauge theory:}
In $\cN=2$ supersymmetric gauge theory, the curve \eqref{class-curve}
is known as the Seiberg-Witten curve \cite{SW-I}, and $\hbar$ is related to the $\Omega$-deformation \cite{Nekrasov}.

\item[\bf 2.] {\bf Chern-Simons theory:}
In $SL(2,\C)$ Chern-Simons theory with a Wilson loop, 
the polynomial $A(x,y)$ is a topological invariant called the $A$-polynomial and plays a role
similar to that of the Seiberg-Witten curve in $\cN=2$ gauge theory \cite{Apol}.
The parameter $\hbar$ is the coupling constant of Chern-Simons theory.

\item[\bf 3.] {\bf Matrix models:}
In matrix models, the curve \eqref{class-curve} is called the spectral curve,
and $\hbar = 1/N$ controls the expansion in (inverse) matrix size \cite{BIPZ}.

\item[\bf 4.] {\bf Topological strings:}
In topological string theory \cite{ADKMV,DHSV}, every curve of the form \eqref{class-curve} defines
a (non-compact) Calabi-Yau 3-fold geometry in which strings propagate,
namely a hypersurface in $(\C^*)^2 \times \C^2$:
\be
A (x,y) \; = \; zw \,.
\ee
The parameter $\hbar$ is the string coupling constant.

\item[\bf 5.] {\bf $\mathcal{D}$-modules:}
There is also a mathematical theory of $\mathcal{D}$-modules \cite{kashiwara,KashiwaraSchapira,Kontsevich},
which studies modules over rings of differential operators,
and in particular operators with properties analogous to those which we expect from $\widehat{A}$.
Some connections of this theory to the above mentioned physics systems were analyzed in \cite{DHSV,DHS,DijkgraafFuji-1}.

\end{itemize}

In all these applications, the primary object of interest is the partition function, $Z(u)$,
or, to be more precise, a collection of functions $Z^{(\alpha)} (u)$ labeled by a choice
of root $v^{(\alpha)} = v^{(\alpha)} (u)$ to the equation \eqref{class-curve}:
\be
Z^{(\alpha)} (u) \; = \; Z (u,v^{(\alpha)} (u)) \,.    \label{Z-alpha}
\ee
The right-hand side of this expression is the partition function $Z(u,v)$,
which is a globally defined function\footnote{To avoid any potential confusion,
we should clarify that even though we write $Z(u,v)$ as a function of $u$ and $v$,
it is meant to be a globally defined function on the Riemann surface (perhaps with a few points removed).
A better way to write it would be $Z(p)$, where $p$ denotes a point on $\cC$,
a notation that we shall use later in section \ref{sec:toprecursion}, {\it cf.} \eqref{param}.
As such $Z(p) = Z(u,v)$ does {\it not} depend on $\alpha$, which labels the choice of sheet in the covering of
the $u$-plane by $\cC$.}  on the Riemann surface \eqref{class-curve}
and which does {\it not} depend on the choice of~$\alpha$.
The existence of such a globally defined partition function is less obvious
in some of the above mentioned applications compared to others.
In our discussion below, we find it more convenient and often more illuminating
to work with $Z(u,v)$ rather than with a collection of functions $Z^{(\alpha)} (u)$.

{}From the viewpoint of quantization, the partition function $Z$ is simply
the wave-function associated to a classical state \eqref{class-curve}.
It obeys a Schr\"odinger-like equation
\be
\widehat{A} Z \; = \; 0 \,,
\label{AhatZ0}
\ee
and has a perturbative expansion of the form
\be
Z \; = \; \exp\left( \frac{1}{\hbar} S_0 
 \,+\,\sum_{n=0}^\infty S_{n+1} \, \hbar^{n} \right)\,.
\label{Zpert}
\ee
The quantum operator $\widehat{A}$ in \eqref{AhatZ0} is precisely the operator
obtained by a quantization of $A(u,v)$ or $A(x,y)$,
and the Schr\"odinger-like equation \eqref{AhatZ0} will be our link
relating its perturbative expansion \eqref{Ahatpert}
to that of the partition function \eqref{Zpert}.

Indeed, recently a number of powerful methods have been developed
that allow to compute perturbative terms $S_n$ in the $\hbar$-expansion.
In particular, insights from matrix models suggest that the perturbative
expansion of the partition function \eqref{Zpert} should be thought of as a large $N$ expansion
of the determinant expectation value in random matrix theory
\be
Z = \Big{\langle}\, \textrm{det}(u-M)\,  \Big{\rangle}  \, .  \label{det-uM}
\ee
This expectation value is computed in some ensemble of matrices $M$ of size $N=\hbar^{-1}$,
with respect to the matrix measure $\mathcal{D} M\, e^{-\textrm{Tr}V(M) /\hbar }$,
where $V(M)$ is a potential of a matrix model.
Then, by exploring the relation between perturbative expansions of $\widehat{A}$ and $Z$,
we argue that having a systematic procedure for computing one is essentially
equivalent to having a similar procedure for the other.
In particular, by shifting the focus to $\widehat{A}$,
we obtain the following universal formula for the first quantum correction $\widehat{A}_1$:
\be
\widehat{A}_1 \; = \;
\frac{1}{2} \left( \frac{\partial_u A}{\partial_v A} \partial_v^2 + \frac{\partial_u T}{T} \partial_v \right) \; A \,,
\label{Aleading}
\ee
expressed in terms of the classical $A$-polynomial and the ``torsion'' $T(u)$
that determines the subleading term in the perturbative
expansion \eqref{Zpert} of the partition function:
\be
S_1 \; = \; - \frac{1}{2} \log T(u) \,.
\label{S1torsion}
\ee
Usually, the torsion is relatively easy to compute, even without detailed knowledge of the higher-order quantum corrections
to \eqref{Ahatpert} or \eqref{Zpert} which typically require more powerful techniques. For instance, in the examples coming from
knot theory the torsion $T(u)$ is a close cousin of the ``classical'' knot invariant called the Alexander polynomial.

Furthermore, it is curious to note that, generically, for curves in $\C^* \times \C^*$ the leading
quantum correction \eqref{Aleading} completely determines the entire quantum operator $\widehat{A}$
when all $\hbar$-corrections can be summed up to powers\footnote{It seems that all polynomials $A(x,y)$
that come from geometry have this property. Why this happens is a mystery.} of $\boxed{q = e^{\hbar}}\,:$
\be
\widehat{A} \; = \; \sum_{(m,n) \in \mathcal{D}}\, a_{m,n}\, q^{c_{m,n}}\, \hat x^m\, \hat y^n \,,
\label{Aqnice}
\ee
in other words, when $\widehat{A}$ can be written as a (Laurent) polynomial in $\hat x$, $\hat y$, and $q$.
Here, $\mathcal{D}$ is a two-dimensional lattice polytope;
in many examples $\mathcal{D}$ is simply the  Newton polygon of $A(x,y)$.
Indeed, the coefficients $a_{m,n}$ are simply the coefficients of the classical polynomial,
$A = \sum a_{m,n} x^m y^n$, which is obtained from \eqref{Aqnice} in the limit $q \to 1$.
On the other hand, the exponents $c_{m,n}$ can be determined by requiring that
\eqref{Aleading} holds for all values of $x$ and $y$ (such that $A(x,y)=0$):
\be
\sum_{(m,n) \in \mathcal{D}}\, a_{m,n}\, c_{m,n}\, x^m y^n \; = \;
\frac{1}{2} \left( \frac{\partial_u A}{\partial_v A} \partial_v^2 + \frac{\partial_u T}{T} \partial_v \right) \; A \,.
\label{Auniversal}
\ee
For curves of low genus this formula takes even a more elementary form \eqref{Agenus0}
which, as we explain, is very convenient for calculations of $\widehat{A}$.
In section \ref{sec:knots} we will illustrate how this works in some simple knot theory examples, and in sections \ref{sec-tetrahedron} and \ref{sec-conifold} in several examples from the topological string theory.

We should emphasize that our desire to illustrate general methods with simple examples is done only
for convenience of the reader and should not be viewed as a limitation of the framework itself,
which is aimed to be completely general and not limited to curves of any particular class.
In fact, starting with simple examples, in this paper we consider quantization of
curves of geometric genus up to 3 and arbitrary arithmetic genus.\footnote{As practitioners
of the topological recursion know very well, it is the latter that determines complexity of a given example.}
Also, as we mentioned earlier, in some cases one can supplement our method based on
the topological recursion with additional shortcuts, which certainly should not be interpreted
as shortcomings of the method itself. For instance, while in examples coming from knots
one can determine \eqref{S1torsion} from the twisted Alexander polynomial, even when this
extra data is not available one can always follow the most direct approach and use
the technique of the topological recursion to systematically compute each term $\widehat A_n$ in \eqref{Ahatpert}.
Depending on the details, explicit computations may be harder in some examples (see {\it e.g.} section \ref{sec:S1}),
but these are merely technical problems and there is nothing conceptual
that prevents computation of $S_n$'s and $\widehat A_n$'s for curves of arbitrary genus.

More importantly, as we illustrate in many examples,
as soon as one knows the first few $A_n$'s, the rest can be determined from \eqref{Auniversal}
or its cousins. It would be interesting to investigate further why this phenomenon happens,
in which examples, and what determines the degree in the perturbative $\hbar$-expansion \eqref{Ahatpert}
that one needs to know in order to determine the rest. We hope to return to these questions in the future work.


\section{Topological recursion versus quantum curves}

In this section, we collect the necessary facts about the perturbative structure of
the partition function \eqref{Zpert} and the Schr\"odinger-like equation \eqref{AhatZ0}
that, when combined together, can tell us how the polynomial $A(u,v)$ or $A(x,y)$ gets quantized,
\be
A \; \leadsto \; \widehat{A} \,.
\label{quantization}
\ee
To the leading order in the $\hbar$-expansion, $\widehat{A}$ is obtained from $A$
simply by replacing $u$ and $v$ by the quantum operators $\hat u$ and $\hat v$.
Then, with the choice of polarization as in \eqref{uvchoice}
the Schr\"odinger-like equation \eqref{AhatZ0} implies the following
leading behavior of the wave-function \eqref{Zpert}:
\bea
S_0
& = & \int v du \qquad\qquad {\rm ~for~curves~in~} \C \times \C \,, \label{Adisk} \\
& = & \int \log y \frac{dx}{x} \,\qquad {\rm ~for~curves~in~} \C^* \times \C^* \,. \nonumber
\eea
In fact, in any approach to quantization
this should be the leading behavior of the semi-classical wave function
associated to the classical state $A=0$.
What about the higher-order terms $S_n$ with $n \ge 1$?

In the introduction we mentioned several recent developments that shed light on the perturbative
(and, in some cases, even non-perturbative) structure of the partition function \eqref{Zpert}.
One of such recent developments is the topological recursion of Eynard-Orantin \cite{eyn-or}
and its extension to curves in $\C^* \times \C^*$ called the ``remodeling conjecture'' \cite{Marino:2006hs,BKMP}.
These techniques are ideally suited for understanding the analytic structure of the quantization \eqref{quantization}.

\subsection{Topological recursion}
\label{sec:toprecursion}

The starting point of the topological recursion \cite{eyn-or} is the choice\footnote{As will be explained
in section \ref{sec:choices}, this choice is related to the choice of polarization.} of a parametrization,
{\it i.e.} a choice of two functions of a local variable $p$,
\be
\left\{\begin{array}{l} u = u(p)  \\
v = v(p)   \end{array}\right.  \label{param}
\ee
where $u(p)$ is assumed to have non-degenerate critical points.
(In particular, for curves of genus zero, both $u(p)$ and $v(p)$ can be rational functions.
We are not going to assume this, however, and, unless noted otherwise,
much of our discussion below applies to curves of arbitrary genus.)
Then, from this data alone one can recursively determine the perturbative coefficients $S_n$
of the partition function \eqref{Zpert} via a systematic procedure that we explain below.

For example, as we already noted in \eqref{Adisk} the leading term $S_0$ is obtained by
integrating a 1-form differential $\phi = v du$ along a path on the curve $A(u,v)=0$.
When expressed in terms of the local coordinate $p$, this integral looks like
\be
S_0 \; = \; \int^p \phi = \int^p v(p) du(p) \,,
\label{def-Phi}
\ee
and sometimes is also referred to as the anti-derivative of $\phi$.
Then, the next-to-the-leading term $S_1$ is determined by the two-point function, or the so-called annulus amplitude.
For a curve $\cC$ of genus zero it can be expressed in terms of the parametrization
data \eqref{param} by the following formula\footnote{Notice, our prescription here and
also in eq. \eqref{Agn-p} differs from that in \cite{DijkgraafFuji-2}. As will be explained below,
these differences are important for overcoming the obstacles in \cite{DijkgraafFuji-2}
and reproducing the ``quantum" $q$-corrections in the quantization of the $A$-polynomial \eqref{quantization}.}
\be
S_1 \; = \; -\frac{1}{2}\log\frac{du}{dp} \,,
\label{A1x}
\ee
whose origin and generalization to curves of arbitrary genus will be discussed in section \ref{sec:S1}.
We recall that, according to \eqref{S1torsion}, the term $S_1$ contains information
about the ``torsion''  $T(u)$ and generically
is all one needs in order to determine
the quantum curve $\widehat{A}$ when it has a nice polynomial form \eqref{Aqnice}.

In a similar manner, the topological recursion of Eynard-Orantin \cite{eyn-or}
can be used to determine all the other higher-order terms $S_n$, $n \ge 2$.
Starting with the parametrization \eqref{param}, one first defines
a set of symmetric degree-$n$ meromorphic differential forms $W^g_n=W^g_n(p_1,p_2,\ldots,p_n)$
on $\cC^n$ via a systematic procedure that we shall review in a moment.
Then, by taking suitable integrals and residues one obtains respectively the desired $S_n$'s,
as well as their ``closed string'' analogs known as the genus-$g$ free energies $F_g$:
\be
\boxed{\phantom{\int} \begin{array}{c@{\qquad}c@{\qquad}c@{\qquad}c@{\qquad}c}
\text{$u(p)$ ~and~ $v(p)$} & \leadsto & \text{$W^g_n$} & \leadsto &  \text{$S_n$ ~and~ $F_g$} \,
\label{conjduality}
\end{array}\phantom{\int}}
\ee
Specifically, motivated by the form of a determinant in (\ref{det-uM}),
or a definition of the Baker-Akhiezer function in \cite{eyn-or,Eynard:2008he,Eynard11104936},
we construct $S_n$'s as the following linear combinations of
the integrated multilinear meromorphic differentials:
\be
S_{n}(p) = \sum_{2g-1+k = n} \frac{1}{k!} \, \underbrace{\int^{p}_{\tilde{p}} \cdots \int^{p}_{\tilde{p}}}_{\text{$k$ times}} \, W^g_k (p'_1,\ldots,p'_k) \,,
\label{Agn-p}
\ee
where each differential form $W^g_k$ of degree $k$ is integrated $k$ times.\footnote{For curves of genus one or higher one should consider more general Baker-Akhiezer function, which in addition includes non-perturbative corrections represented by certain $\theta$-functions \cite{Eynard:2008he}. As the examples which we consider concern mostly curves of genus zero, we do not analyse such corrections explicitly.} The base point of integration $\tilde{p}$ is chosen such that $u(\tilde{p})\to \infty$ \cite{Eynard11104936}. 
In turn, the multilinear differentials $W^g_n$ are obtained by
taking certain residues around critical points of the ``Morse function'' $u(p)$,
{\it i.e.} solutions to the equations
\be
du(p)\vert_{p^*_i}=0 \qquad \Leftrightarrow \qquad  \partial_v A \vert_{p^*_i}=0 \,,
\label{branchpoint}
\ee
where the standard shorthand notation $\partial_v \equiv \frac{\partial}{\partial v}$ is used.
Following \cite{eyn-or}, we shall refer to these points as the ``branch points''
of the curve $\cC$ in parametrization \eqref{param}.
We assume that all branch points are simple (or, as sometimes referred to, regular), i.e. 
for each point $p$ in the neighborhood of a branch point $p^*_i$ there is a unique,
conjugate point $\bar{p}$, such that
\be
u(p) \; = \; u(\bar{p}) \,.
\label{conjugate}
\ee


The next essential ingredient for the topological recursion is the differential 1-form\footnote{For reasons
that will become clear later, we choose a sign opposite to the conventions of \cite{eyn-or}.} called the ``vertex'':
\bea
\omega(p)
& = & \big( v(\bar{p}) - v(p) \big) du(p) \,\,\,\,\,\,\,\,\qquad\qquad {\rm ~for~curves~in~} \C \times \C \,,   \label{def-omega} \\
& = & \big( \log y(\bar{p}) - \log y(p) \big) \frac{dx(p)}{x(p)} \qquad {\rm ~for~curves~in~} \C^* \times \C^* \,, \nonumber
\eea
and the 2-form $B(p,q)$ known as the Bergman kernel.
The Bergman kernel $B(p,q)$ is defined as the unique meromorphic differential with exactly one pole,
which is a double pole at $p=q$ with no residue, and with vanishing integral over $A_I$-cycles $\oint_{A_I} B(p,q)=0$
(in a canonical basis of cycles $(A_I,B^I)$ for $\cC$).
Thus, for curves of genus zero the Bergman kernel takes a particularly simple form
\be
B(p,q) \; = \; \frac{dp\,dq}{(p-q)^2} \,,
\label{Bergman}
\ee
and its form for curves of higher genus is presented in section \ref{sec:S1}.
A closely related quantity is a 1-form, defined in a neighborhood of a branch point $q^*_i$
$$
dE_q(p) \; = \; \frac{1}{2} \int_q^{\bar{q}} B(\xi,p) \,.
$$
Finally, the last important ingredient is the recursion kernel $K(q,p)$,
\be
K(q,p) \; = \; \frac{dE_q(p)}{\omega(q)} \,.
\label{kernel}
\ee

Having defined the above ingredients we can present the recursion itself.
When expressed in variables $(u,v)$, the recursion has the same form for curves in $\C \times \C$
as it does for curves in $\C^* \times \C^*$.
It determines higher-degree meromorphic differentials $W^g_n (p_1, \ldots, p_n)$ from those of lower degree.
The initial data for the recursion are one- and two-point correlators of genus zero,
the former vanishing by definition and the latter given by the Bergman kernel:
\bea
{\raisebox{-.3cm}{\includegraphics[width=0.7cm]{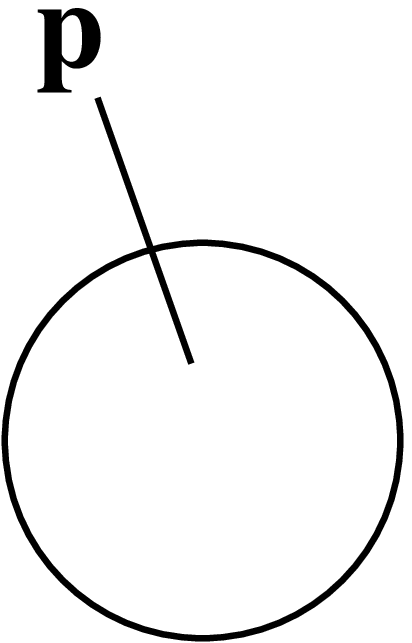}}\,}
& : \qquad &
W^0_1(p) \; = \; 0 \,,  \label{W01} \\
{\raisebox{-.3cm}{\includegraphics[width=0.9cm]{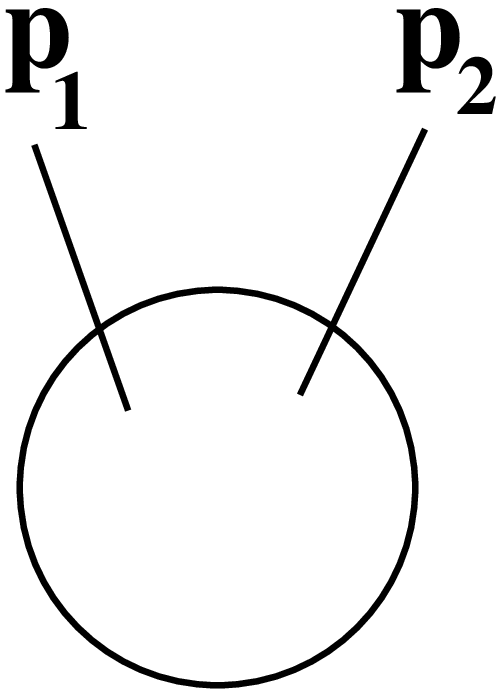}}\,}
& : \qquad &
W^0_2(p_1,p_2) \; = \; B(p_1,p_2) \,.  \label{W02}
\eea
It is also understood that $W^{g<0}_n=0$.

\bigskip
\begin{figure}[ht]
\centering
\includegraphics[width=5.0in]{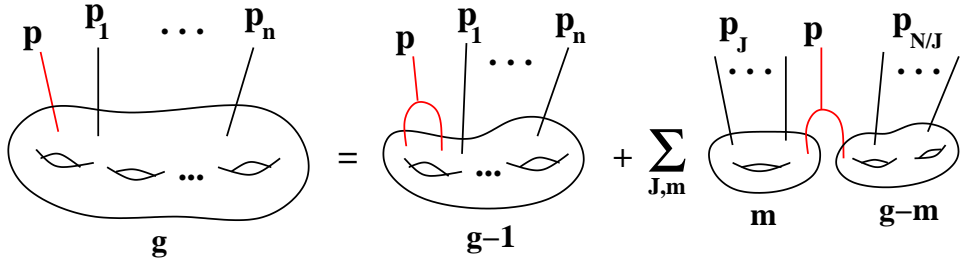}
\caption{A graphical representation of the Eynard-Orantin topological recursion.}
\label{fig:recursion}
\end{figure}

The other differentials are defined recursively as follows.
For a set of indices $J$ denote $\vec{p}_J = \{p_i  \}_{i\in J}$.
Then, for $N=\{1,\ldots,n\}$ and the corresponding set of points $\vec{p}_N=\{p_1,\ldots,p_n\}$ define
\bea
{\raisebox{-.6cm}{\includegraphics[width=2.0cm]{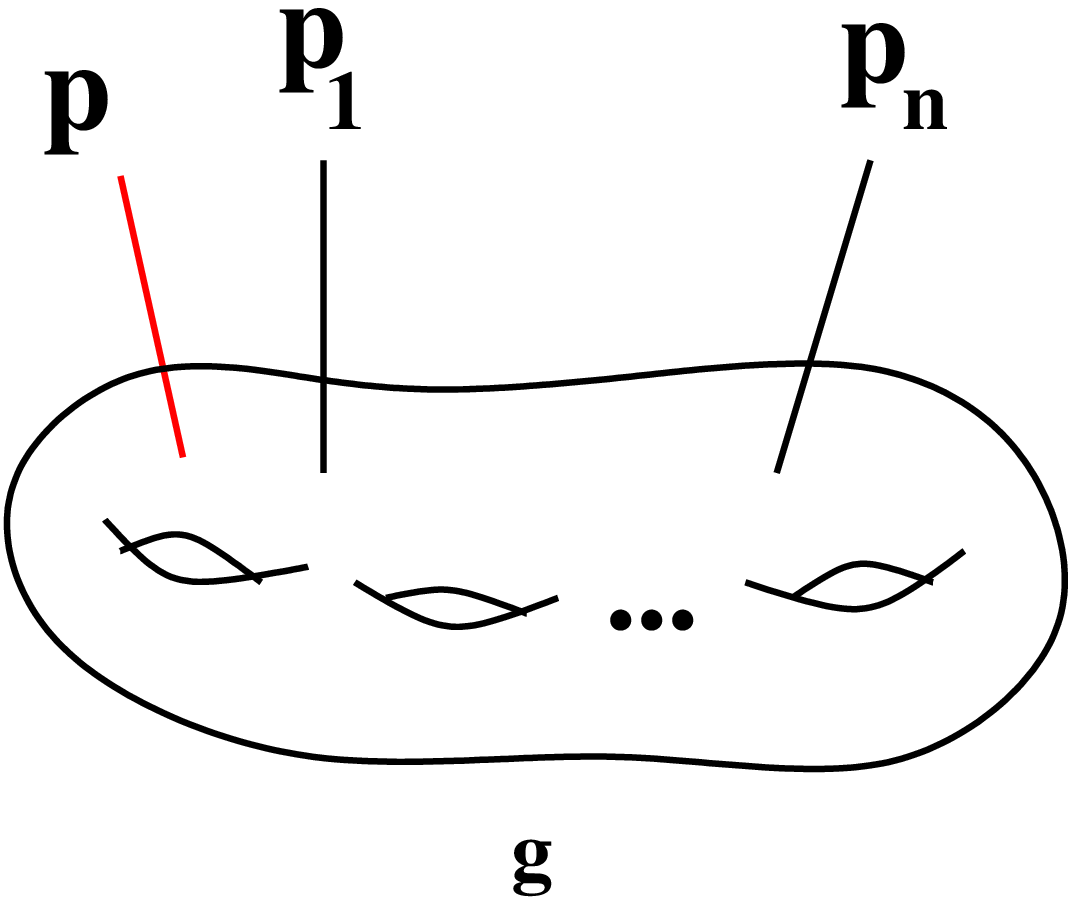}}\,} : \quad
W^g_{n+1}(p,\vec{p}_N)
& = & \sum_{q^*_i}\textrm{Res}_{q\to q^*_i}  K(q,p)  \Big(  W^{g-1}_{n+2}(q,\bar{q},\vec{p}_N)  +  \nonumber \\
& & + \sum_{m=0}^g\sum_{J\subset N} W^m_{|J|+1}(q,\vec{p}_J) W^{g-m}_{n-|J|+1}(\bar{q},\vec{p}_{N/J}  \Big),  \label{top-recursion}
\eea
where $\sum_{J\subset N}$ denotes a sum over all subsets $J$ of $N$, {\it cf.} Figure \ref{fig:recursion}.
These correlators have many interesting properties.
For example, any $W^g_n(p_1,\ldots,p_n)$ is a symmetric function of $p_i$.
Furthermore, apart from the special case of $g=0$ and $n=2$,
the poles of $W^g_n(p_1,\ldots,p_n)$ in variables $p_i$ appear only at the branch points.
In addition, the $A_I$-cycle integrals with respect to any $p_i$ vanish, $\oint_{p_i\in A_I} W^g_n(p_1,\ldots,p_n)=0$.
For a detailed discussion of these and many other features of $W^g_n$ see \cite{eyn-or}.

Let us briefly illustrate how the recursion procedure works.
First, from the recursion kernel \eqref{kernel} and from the Bergman kernel \eqref{W02}
one finds the genus-1 one-point correlator
\be
W^1_1(p) \; = \; \sum_{q^*_i}\textrm{Res}_{q\to q^*_i}  K(q,p) W^0_2(q,\bar{q}) \,.   \label{W11}
\ee
Then, the following series (with $g+n=3$) is determined
\bea
W^0_3(p,p_1,p_2) & = & {\, \raisebox{-.3cm}{\includegraphics[width=0.9cm]{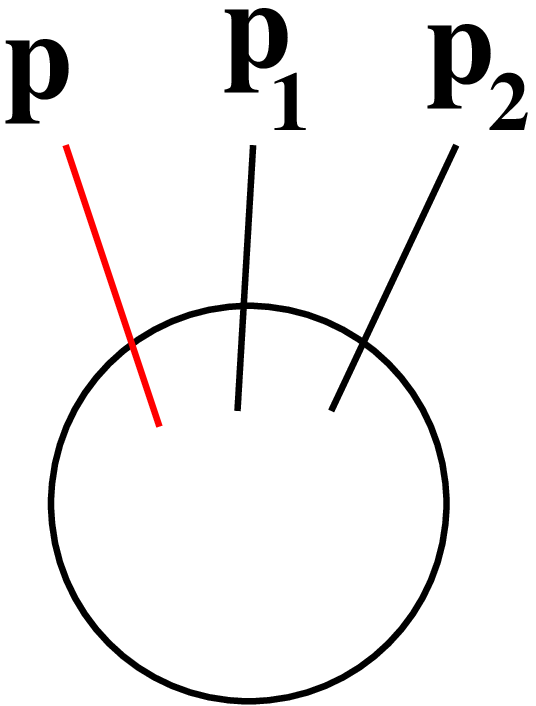}}\,} \label{W03} \\
& = & \sum_{q^*_i}\textrm{Res}_{q\to q^*_i}  K(q,p) \Big( W^0_2(q,p_1) W^0_2(\bar{q},p_2) + W^0_2(\bar{q}, p_1) W^0_2(q,p_2)  \Big) \,, \nonumber \\
W^1_2(p,p_1) & = & {\, \raisebox{-.3cm}{\includegraphics[width=0.8cm]{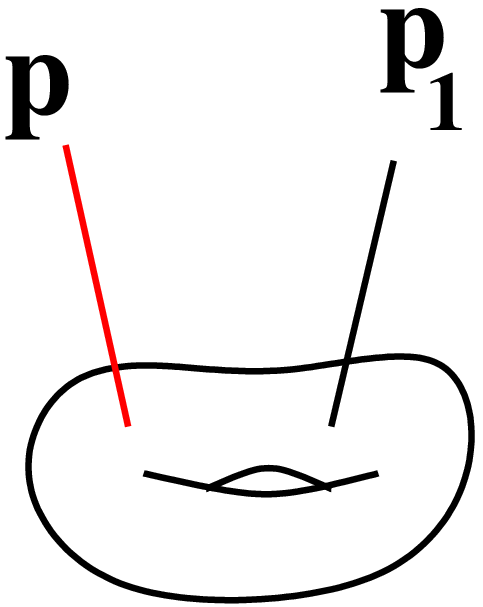}}\,}
= \sum_{q^*_i}\textrm{Res}_{q\to q^*_i}  K(q,p) \Big( W^0_3(q,\bar{q}, p_1) + 2 W^1_1(q) W^0_2(\bar{q}, p_1)  \Big) \,,  \label{W12} \\
W^2_1(p) & = & {\, \raisebox{-.3cm}{\includegraphics[width=1.1cm]{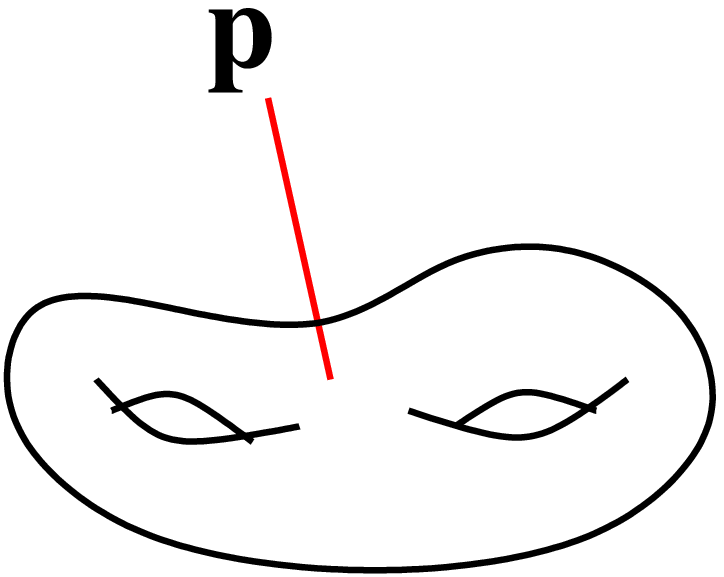}}\,}
= \sum_{q^*_i}\textrm{Res}_{q\to q^*_i}  K(q,p) \Big( W^1_2(q,\bar{q}) + W^1_1(q) W^1_1(\bar{q}) \Big) \,.  \label{W21}
\eea
Next, one finds a series $W^0_4, W^1_3, W^2_2, W^3_1$ with $g+n=4$, and so on.
In the end, from each such series one can determine one more $S_n$ using \eqref{Agn-p}.
For example, as will be discussed in section \ref{sec:S1}, $S_1$ is obtained by
integrating the Bergman kernel:
\be
S_1(p) \; = \; \frac{1}{2} \lim_{p_1 \to p_2=p} \int
\left( B(p_1,p_2) - \frac{du (p_1) \, du (p_2)}{(u(p_1) - u(p_2))^2} \right) \,,
\label{Aannulus}
\ee
and for curves of genus zero this formula reproduces the expression \eqref{A1x} proposed earlier.
At the next step, from the series of the multilinear differentials \eqref{W03} - \eqref{W21}
one finds the next term in the perturbative series \eqref{Zpert}:
\be
S_2(p) \; = \; \int^p_{\tilde{p}} W^1_1(p_1) + \frac{1}{3!} \int^p_{\tilde{p}} \int^p_{\tilde{p}} \int^p_{\tilde{p}} W^0_3 (p_1,p_2,p_3) \,,
\label{A2x}
\ee
and so on. As noted before, the base point of integration $\tilde{p}$ is chosen such that $u(\tilde{p})\to \infty$ \cite{Eynard11104936}. 

While not of our immediate concern in this paper,
for completeness we also recall a definition of genus-$g$ free energies $F_g$. For $g\geq 2$ they come\footnote{Notice, compared to the conventions of \cite{eyn-or} we introduce an extra minus sign in our definition of $F_g$ in order to account for the sign of $W^g_1$ originating from the sign in (\ref{def-omega}).} from the corresponding $W^g_1$:
\be
F_g \; = \; \frac{1}{2g-2} \sum_{q^*_i} \textrm{Res}_{q\to q^*_i} S_0 (q) W^g_1(q) \,,     \label{Fg}
\ee
where $S_0 (q)=\int^q v(p) du(p)$, while $F_0$ and $F_1$ are defined independently in a more intricate way presented in \cite{eyn-or}.
Among various interesting properties of $F_g$ the most important one is their invariance under symplectic transformations of the spectral curve.

Finally, since the relation between $S_n$ and $W^g_k$ will be crucial for computing $\widehat{A}$ from the classical curve $A=0$ and its parametrization, let us briefly explain our motivation behind \eqref{Agn-p}. Recall, that the correlators $W^g_k (p_1,\ldots,p_k)$ in (\ref{Agn-p}) were originally introduced \cite{eyn-or} in a way which generalizes and, when an underlying matrix model exists, reproduces connected contributions to the matrix model expectation value
$$
\Big{\langle} \textrm{Tr} \Big( \frac{1}{u(p_1) - M} \Big) \cdots \textrm{Tr} \Big( \frac{1}{u(p_k) - M} \Big) \Big{\rangle}_{{\rm conn}}  = \sum_{g=0}^{\infty} \hbar^{2g-2+k} \frac{W^g_k (p_1,\ldots,p_k)}{du(p_1) \ldots du(p_k)}
$$
in an ensemble of matrices $M$ of size $N=\hbar^{-1}$.\footnote{Strictly speaking, this equation holds for $k>2$ and there are some corrections to the lowest order terms with $k=1$ and $k=2$ \cite{eyn-or}.} Integrating both sides with respect to all variables and then setting $p_1=\ldots=p_k = p$, we get
$$
\Big{\langle} \Big( \textrm{Tr} \log \big(u(p)-M \big) \Big)^k \Big{\rangle}_{{\rm conn}}
 =  \sum_{g=0}^{\infty} \hbar^{2g-2+k}  \int^p\cdots \int^p  W^g_k (p_1',\ldots,p_k').
$$
Dividing both sides by $k!$ and summing over $k$ we get
$$
\Big{\langle}\, \textrm{det}(u-M)\,  \Big{\rangle}_{{\rm conn}} = \sum_{n=0}^{\infty} \hbar^{n-1} S_n(p),
$$
with $S_n(p)$ defined in (\ref{Agn-p}). Whereas the left hand side represents the connected expectation value, the right hand side plays the role of the free energy, so that
$$
Z = \Big{\langle}\, \textrm{det}(u-M)\,  \Big{\rangle} = e^{ \frac{1}{\hbar} \sum_{g=0}^{\infty} \hbar^{n} S_n(p) }.
$$
This result is in agreement with (\ref{Zpert}) and (\ref{det-uM}) and provides the motivation for the definition (\ref{Agn-p}). From the matrix model point of view, the free energies $F_g$ defined in (\ref{Fg}) encode the total partition function
\be
\langle 1 \rangle = \int \mathcal{D}M e^{-\frac{1}{\hbar}\textrm{Tr} V(M)} = e^{\sum_{g=0}^{\infty} \hbar^{2g-2} F_g} \,.
\label{Zclosed}
\ee
{}From a string theory viewpoint, this partition function would correspond to closed string amplitudes.
In fact, in many instances relevant to Seiberg-Witten theory or topological strings, matrix models
which encode corresponding partition functions \eqref{Zclosed} have been explicitly constructed
in \cite{Aganagic:2002wv,Eynard:2008mt,Klemm:2008yu,Sulkowski:2009br,Eynard:2010dh,Eynard:2010vd,OSV}.


\subsection{Quantum curves and differential hierarchies}
\label{sec:hierarchy}

Our next goal is to compare the results of the topological recursion to the structure of the ``quantum curve''
\be
\widehat{A} \; \simeq \; 0 \,,
\label{AhatnoZ}
\ee
where we used a shorthand notation ``$\simeq$'' to write 
\eqref{AhatZ0} in a form that makes a connection with its classical limit $A(x,y)=0$ manifest, {\it cf.} \cite{Tudor}.
In general, the Schr\"odinger-like equation \eqref{AhatZ0} and its abbreviated form \eqref{AhatnoZ}
is either a $q$-difference equation (for curves in $\C^* \times \C^*$)
or an ordinary differential equation (for curves in $\C \times \C$).
In either case, we need to write it as a power series in $\hbar$,
which was the expansion parameter in the topological recursion.

In practice, one needs to substitute the perturbative expansions \eqref{Ahatpert} and \eqref{Zpert}
into the Schr\"odinger-like equation \eqref{AhatZ0},
\be
\left( \widehat{A}_0 + \hbar \widehat{A}_1 + \hbar^2 \widehat{A}_2 + \ldots \right)
\exp \left( \frac{1}{\hbar} \sum_{n=0}^\infty S_{n} \, \hbar^{n} \right) \; = \; 0 \,,   \label{AhatZ}
\ee
and collect all terms of the same order in $\hbar$-expansion.
This requires some algebra (see \cite{DGLZ} and appendix \ref{app:hierarchy}), but after the dust settles
one finds\footnote{Once again, we point out that,
when expressed in terms of variables $u$ and $v$, most of our formulas have the same form
on {\it any} complex symplectic 2-fold with the holomorphic symplectic 2-form \eqref{sympform}.
In particular, the hierarchy of differential equations \eqref{hierarchy-SA} written in variables $(u,v)$
looks identical for curves in $\C\times\C$ and in $\C^*\times\C^*$.
Of course, the reason is simple: it is not the algebraic structure, but, rather, the symplectic structure that matters in the quantization problem.
For this reason, throughout the paper we write most of our general formulas in variables $(u,v)$ with understanding that, unless noted otherwise,
they apply to curves in arbitrary complex symplectic 2-fold with the holomorphic symplectic 2-form \eqref{sympform}.}
a nice hierarchy of loop equations
\be
\boxed{\phantom{\int}
\sum_{r=0}^n \D_{r} A_{n-r} \; = \; 0 \,,
\label{hierarchy-SA}    \phantom{\int}}
\ee
expressed in terms of symbols $A_{n-r}$ of the operators $\widehat{A}_{n-r}$
and in terms of differential operators $\D_r$.
Specifically, each $\D_r$ is a differential operator of degree $2r$; it can be
written as a degree-$2r$ polynomial in $\partial_v \equiv \frac{\partial}{\partial v}$,
whose coefficients are polynomial expressions in functions $S_k (u)$ and their derivatives.
For example, the first few differential operators look like
\begin{subequations}\label{ddd}
\bea
\D_0 & = & 1 \,, \\
\D_1 & = & \frac{S''_0}{2}  \partial_v^2 + S'_1 \partial_v \,,  \\
\D_2 & = & \frac{(S''_0)^2}{8} \partial_v^4 +  \frac{1}{6}\big(S'''_0 + 3S''_0 S'_1 \big) \partial_v^3
+ \frac{1}{2}\big(S''_1 + (S'_1)^2 \big) \partial_v^2 + S'_2 \partial_v \,, \\
& \vdots & \nonumber
\eea
\end{subequations}
and yield the corresponding equations, at each order $\hbar^n$ in \eqref{hierarchy-SA}:
\bea
\hbar^0 & : \qquad & A \; = \; 0 \,, \label{Aclass} \\
\hbar^1 & : \qquad & \Big(\frac{S''_0}{2}\partial_v^2 + S'_1 \partial_v \Big) A + A_1 \; = \; 0 \,, \label{hier-eq2v} \\
& & \quad \vdots \nonumber \\
\hbar^n & : \qquad & \D_n A + \D_{n-1} A_1 + \ldots + A_n \; = \; 0 \,, \label{hier-SA-Cstar} \\
& & \quad \vdots \nonumber
\eea
The first equation is equivalent to the classical curve equation \eqref{class-curve},
provided $S_0' \equiv \frac{dS_0}{du} = v$ which, in turn, leads to the expression \eqref{Adisk} for $S_0 (u)$.
The second equation \eqref{hier-eq2v} is also familiar from \eqref{Aleading} and \eqref{Auniversal},
where the second order differential operator $\D_1$ acting on $A_0 \equiv A$ was expressed in terms of the ``torsion'' $T(u)$.
If we know the partition function $Z$, then, at each order $\hbar^n$, the above equations uniquely determine the correction $\widehat{A}_n$; or \emph{vice versa}: from the knowledge of the total $\widehat{A}$, at each order order $\hbar^n$, we can
determine $S_n$ (up to an irrelevant normalization constant).

More generally, the operators $\D_r$ are defined via the generating function
\be
\sum_{r=0}^{\infty} \hbar^r \D_r \; = \;
\exp \left( \sum_{n=1}^{\infty} \hbar^n \frak{d}_n \right) \,,
\label{Sigma-def}
\ee
where
\be
\frak{d}_n \; = \; \sum_{r=1}^{n+1} \frac{S_{n+1-r}^{(r)}}{r!} (\partial_v)^r \,.
\ee
For example, the explicit expressions for 
small values of $n$
\bea
\frak{d}_1 & = & \frac{1}{2} S''_0 \partial_v^2 + S'_1 \partial_v \,, \nonumber \\
\frak{d}_2 & = & \frac{1}{6}S'''_0 \partial_v^3 + \frac{1}{2} S''_1 \partial_v^2 + S'_2 \partial_v \,, \nonumber \\
\frak{d}_3 & = & \frac{1}{4!}S^{(4)}_0 \partial_v^4 +\frac{1}{3!}S'''_1 \partial_v^3 + \frac{1}{2} S''_2 \partial_v^2 + S'_3 \partial_v \,, \nonumber
\eea
lead to the formulas (\ref{ddd}). More details and a derivation of the above hierarchy
are given in appendix \ref{app:hierarchy}.

Our goal in the rest of the paper is to combine
the steps in sections \ref{sec:toprecursion} and \ref{sec:hierarchy}
into a single technique that can produce a quantum operator $\widehat{A}$
starting with a parametrization of the classical curve \eqref{class-curve}, much as in the topological recursion:
\be
\begin{array}{c@{\qquad}c@{\qquad}c}
\text{$u(p)$ ~and~ $v(p)$} & \leadsto & \widehat{A} \,.
\label{uvgivesA}
\end{array}
\ee
Basically, one can use the output of \eqref{conjduality} as an input for
\eqref{Aclass}-\eqref{hier-SA-Cstar} (written more compactly in \eqref{hierarchy-SA})
to produce a perturbative expansion \eqref{Ahatpert}.


\subsection{Parametrizations and polarizations}
\label{sec:choices}

The quantization procedure \eqref{quantization} on one hand,
and the topological recursion \eqref{conjduality} on the other
come with certain inherent ambiguities which are not unrelated.

In quantization, one needs to split the coordinates on the phase space into
``configuration space coordinates'' and ``conjugate momenta.'' This choice, called the choice
of polarization, means that one needs to pick a foliation of the phase space by
Lagrangian submanifolds parametrized by a maximal set of mutually commuting ``coordinates''
(with the remaining variables understood as their conjugate momenta).
In the problem at hand, the (complex) phase space is 2-dimensional, with the symplectic form \eqref{sympform},
\be
\omega \; = \; \frac{i}{\hbar} du \wedge dv \,,
\ee
so that the ambiguity associated with the choice of polarization is described by
one functional degree of freedom, say, a choice of function $f(u,v)$ that one regards as a ``coordinate.''
Thus, in most of the present paper we make a natural\footnote{in most applications} choice \eqref{uvchoice}
treating $u$ as the ``coordinate'' and $v$ as the momentum.
Any other choice is related to this one by a canonical transformation
\be
v = \frac{\partial \mathcal{W}}{\partial u}  \,, \qquad V = - \frac{\partial \mathcal{W}}{\partial U}
\label{canonicaltransform}
\ee
that depends on a single function $\mathcal{W}(u,U)$.
By definition, the transformation $(u,v) \mapsto (U,V)$ preserves the symplectic form $\omega$.
For example, $U = v$ and $V = -u$ corresponds to $\mathcal{W}(u,U)=uU$.

Similarly, as we reviewed in section \ref{sec:toprecursion}, the ambiguity in the topological recursion
is also described by a single function $u(p)$ that enters the choice of parametrization \eqref{param}.
(The functional dependence of $v(p)$ is then determined, up to a discrete action
of the Galois group permuting branches $v^{(\alpha)}$, by the condition $A(u,v)=0$.)
Indeed, starting with different parametrizations of the same classical curve \eqref{class-curve}
and following \eqref{uvgivesA} one arrives at different expressions for $\widehat{A}$.
To make a contact with the choice of polarization, let us point out that part of its ambiguity is
already fixed in the topological recursion (since $u(p)$ is a function of a single variable, whereas
$\mathcal{W}(u,U)$ in \eqref{canonicaltransform} is a function of two variables).
However, a transformation from $u(p)$ to $U(p)$ can be understood as a particular symplectic
transformation $(u,v) \mapsto (U,V)$, such that $U = f(u)$ and $V = v / f'(u)$.
For example, a simple choice of $f(u) = u + c$ with a constant $c$ corresponds to
\be
U \; = \; u + c \qquad , \qquad V = v \,,
\ee
and does not affect $\widehat{A}$.
On the other hand, a similar ``shift transformation'' of the momentum~$v$,
\be
\hat v = \hbar \partial_u ~ \to ~ \hat v = \hbar \partial_u + c \hbar
\label{vshift}
\ee
is equivalent to $Z(u) \to e^{cu} Z(u)$ and, therefore, transforms the quantum operator $\widehat{A}$ as
\be
\widehat{A} (\hat x, \hat y) ~ \to ~ \widehat{A} (\hat x, q^c \hat y) \,.
\label{vshiftonA}
\ee
This transformation plays an important role in our applications since it controls
a (somewhat ambiguous) constant term in $S_1'$.

We also note that, with the choice of uniformization (\ref{param}) and in the polarization
where $p$ is the ``coordinate'' the quantum curve factorizes to the leading order in $\hbar$
\be
\widehat{A} = \prod_{\alpha} \big(\hbar \partial_p + f^{(\alpha)}(p) \big) +\mathcal{O}(\hbar) \,.
\label{Ahat-factor}
\ee
Then, to the leading order in $\hbar$, various branches of the partition function (\ref{Z-alpha}) are annihilated by the first order operators $(\hbar \partial_p + f^{(\alpha)}(p) )$, so that
\be
Z^{(\alpha)} = e^{-\frac{1}{\hbar} \int f^{(\alpha)}(p) dp} \Big( 1 + \mathcal{O}(\hbar) \Big) \,.
\nonumber 
\ee


\subsection{Relation to algebraic K-theory}
\label{sec:ktheory}

Now we come to a very important point, which could already have been emphasized much earlier in the paper:
\begin{center}
{\it Not every curve $\cC$ defined by the zero-locus of a polynomial $A$ is ``quantizable''!}
\end{center}
Namely, one can always produce a non-commutative deformation of the ring of functions
on $\C \times \C$ or $\C^* \times \C^*$, which obeys \eqref{uvcomm} with $\hbar$ as a formal parameter and,
therefore, at least formally gives \eqref{quantization}. However, in physics, one is usually interested
in the actual (not formal) deformation of the algebra of functions with a parameter $\hbar$ and, furthermore,
it is important to know whether a state associated with a particular Lagrangian submanifold in the classical phase space
exists in the Hilbert space of the quantum theory.

In the present case, this means that not every Lagrangian submanifold defined by the zero locus of $A(x,y)$
corresponds to an actual state in the Hilbert space of the quantum theory; the ones which do we call\footnote{Notice,
{\it a priori} this definition of ``quantizability'' has nothing to do with the nice property \eqref{Aqnice}
exhibited by many quantum operators $\widehat{A}$ that come from physical problems; one can imagine a perfectly
quantizable polynomial $A(x,y)$ in the sense described here, for which the quantum corrections \eqref{Ahatpert}
can {\it not} be summed up into a finite polynomial of $x$, $y$, and $q$.
We plan to elucidate the relation between these two properties in the future work.} ``quantizable.''
Specifically, whether the solution to the quantization problem exists or not depends on the complex
structure\footnote{At first, this may seem a little surprising, because the quantization problem is about symplectic geometry
and not about complex geometry of $\cC$. (Figuratively speaking, quantization aims to replace all classical objects
in symplectic geometry by the corresponding quantum analogs.)
However, our ``phase space,'' be it $\C \times \C$ or $\C^* \times \C^*$, is very special in a sense that
it comes equipped with a whole $\mathbb{C}{\bf P}^1$ worth of complex and symplectic structures,
so that each aspect of the geometry can be looked at in several different ways, depending on which complex or symplectic structure we choose.
This hyper-K\"ahler nature of our geometry is responsible, for example, for the fact that a curve $\cC$ ``appears'' to be holomorphic (or algebraic).
We put the word ``appears'' in quotes because this property of $\cC$ is merely an accident, caused by the hyper-K\"ahler structure
on the ambient space, and is completely irrelevant from the viewpoint of quantization. What is important to the quantization
problem is that $\cC$ is Lagrangian with respect to the symplectic form \eqref{sympform}.\label{hyperK}}
of the curve $\cC$, {\it i.e.} on the coefficients of the polynomial $A(x,y)$ that defines it.

Following \cite{Apol}, we explain this important point in a simple example of, say, the figure-8 knot.
Relegating further details to the next section, let us take a quick look at the classical curve
\be
\cC ~: \qquad
x^4 - (1 - x^2 - 2x^4 - x^6 + x^8) y + x^4 y^2 = 0
\label{curve41}
\ee
defined by the zero locus of the $A$-polynomial of the figure-8 knot (see table \ref{table}).
This polynomial equation has a number of special properties, including integrality of coefficients,
symmetries (with respect to $x \to 1/x$ and $y \to 1/y$), and so on.
More importantly, the classical curve \eqref{curve41} is quantizable.

Preserving most of the nice properties of \eqref{curve41} we can make a tiny change to the polynomial
$A(x,y)$ to obtain a close cousin of $\cC$:
\be
\cC' ~: \qquad
x^4 - (x^{-2} - x^2 - 2x^4 - x^6 + x^{10}) y + x^4 y^2 = 0
\label{curve41fake}
\ee
To a naked eye, there is almost no difference between the curves $\cC$ and $\cC'$;
indeed, every obvious property of one is manifest in the other and vice versa. Nevertheless, the curve \eqref{curve41} defined by the true
$A$-polynomial of the figure-8 knot is quantizable, whereas the counterfeit \eqref{curve41fake} is not. Why?

The reason, as explained in \cite{Apol} for Chern-Simons theory and in \cite{OVV,OSV,Evslin,MPP} for topological strings,
is that all periods of the 1-form ${\rm Im} \, \phi$ must vanish
\be
\oint_{\gamma} \Big( \log |x| d ({\rm arg} \, y) - \log |y| d ({\rm arg} \, x) \Big) \; = \; 0 \,,
\label{qcond0}
\ee
and, furthermore, the periods of the 1-form ${\rm Re} \, \phi$ should be integer (or, at least, rational)
multiples of $2 \pi i$ or, equivalently,
\be
\frac{1}{4 \pi^2} \oint_{\gamma} \Big( \log |x| d \log |y| + ({\rm arg} \, y) d ({\rm arg} \, x) \Big) \; \in \; \mathbb{Q}
\label{qcondQ}
\ee
for all closed paths $\gamma$ on the curve $\cC$, from which the zeros or poles of $x$ and $y$ are removed.
Indeed, these two conditions guarantee that
$Z = \exp \big( \frac{1}{\hbar}S_0 + \ldots \big) = \exp \big( \frac{1}{\hbar} \int^p \phi + \ldots \big)$
is well-defined and, therefore, they represent the necessary conditions for $A(x,y)=0$ to be
quantizable.\footnote{Notice, various choices discussed in section \ref{sec:choices} lead to expressions for $\phi$
which differ by (non-holomorphic) exact terms. For more details on change of polarization see {\it e.g.} \cite{GMtorsion}.}
It is not difficult to verify that these conditions are met for the curve \eqref{curve41} but not for the curve \eqref{curve41fake}.

Notice, the constraints \eqref{qcond0}--\eqref{qcondQ} are especially severe for curves of high genus.
Moreover, these constraints have an elegant interpretation\footnote{We thank D.~Zagier for helpful discussions on this point.} in terms of algebraic K-theory and the Bloch group of~$\overline{\mathbb{Q}}$.
To explain where this beautiful connection comes from, we start with the observation that
the left-hand side of \eqref{qcond0} is the image of the symbol $\{ x,y \} \in K_2 (\cC)$ under
the regulator map\footnote{defined by Beilinson \cite{Beilinson} after Bloch \cite{Bloch}}
\bea
r ~:~ K_2 (\cC) & \to & H^1 (\cC, \R) \label{regmap} \\
\{ x,y \} & \mapsto & \eta (x,y) \nonumber
\eea
evaluated on the homology class of a closed path $\gamma$ that avoids all zeros and poles of $x$ and~$y$,
see \cite{DJZagier} for a nice exposition.
Indeed, the left-hand side of \eqref{qcond0} is the integral of the real differential 1-form on $\cC$
(with zeros and poles of $x$ and $y$ excluded),
\be
\eta (x,y) \; = \; \log |x| d ({\rm arg} \, y) - \log |y| d ({\rm arg} \, x) \,,
\ee
which, by definition, is anti-symmetric,
\be
\eta (y,x) = - \eta (x,y) \,,
\ee
obeys the ``Leibniz rule,''
\be
\eta (x_1 x_2, y) = \eta (x_1, y) + \eta (x_2, y) \,,
\ee
and, more importantly, is closed
\be
d \eta (x,y) = {\rm Im} \left( \frac{dx}{x} \wedge \frac{dy}{y} \right) = 0 \,.
\label{closedeta}
\ee
For curves, the latter condition is almost trivial and immediately follows from dimensional considerations,
which is another manifestation of the ``accidental'' extra structure discussed in the footnote \ref{hyperK}.
In higher dimensions, however, the condition \eqref{closedeta} is very non-trivial and holds precisely when $\cC$
is Lagrangian with respect to (real / imaginary part of) the symplectic form~\eqref{sympform}.

We have learnt that the differential 1-form $\eta (x,y)$ is closed. However, to meet the condition \eqref{qcond0}
and, ultimately, to reformulate this condition in terms of algebraic K-theory we actually want $\eta (x,y)$ to be exact.
In order to understand when this happens, it is important
to describe $\eta (x,y)$ near those points on $\cC$ where rational functions $x,y \in \C (\cC)^*$ have zeros or poles.
Let $p$ be one of such points and let ${\rm ord}_p (x)$ (resp. ${\rm ord}_p (y)$) be the order of $x$ (resp. $y$) at $p$.
Then, we have
\be
\frac{1}{2\pi} \oint \eta (x,y) \; = \; \log |(x,y)_p|
\label{reseta}
\ee
where the integral is over a small circle centered at $p$ and
\be
(x,y)_p = (-1)^{{\rm ord}_p (x)\, {\rm ord}_p (y)} \frac{x^{{\rm ord}_p (y)}}{y^{{\rm ord}_p (x)}} \Big{\vert}_p
\label{tamesymb}
\ee
is the {\it tame symbol} at $p \in \cC$.

One general condition that guarantees vanishing of \eqref{reseta}
is to have $\{ x,y \} = 0$ in $K_2 (\C (\cC)) \otimes \Q$.
Then, all tame symbols \eqref{tamesymb} are automatically torsion and $\eta (x,y)$ is actually exact, see {\it e.g.} \cite{LiWang}.
Motivated by this, we propose the following criterion for quantizability:
\be
\boxed{\phantom{\int^1}
\cC~ \text{is quantizable} \qquad \Longleftrightarrow \qquad \{ x,y \} \in K_2 (\C (\cC)) ~\text{is a torsion class}
\phantom{\int^1}}
\label{quantcritK2}
\ee
This criterion is equivalent \cite{Fernando} to having
\be
x \wedge y \; = \; \sum_i r_i z_i \wedge (1 - z_i) \qquad \qquad {\rm in~} \wedge^2 (\C (\cC)^*) \otimes \Q
\label{qcondzz}
\ee
for some $z_i \in \C (\cC)^*$ and $r_i \in \Q$.
When this happens, one can write
\be
\eta (x,y) = d \left( \sum_i r_i D (z_i) \right) = dD \left( \sum_i r_i [z_i] \right)
\label{etaviaD}
\ee
in terms of the Bloch-Wigner dilogarithm function,
\be
D(z) \; := \; \log |z| {\rm arg} (1-z) + {\rm Im} ({\rm Li}_2 (z)) \,,
\ee
which obeys the famous 5-term relation
\be
D(x) + D(y) + D(1 - xy) + D \Big( \frac{1-x}{1-xy} \Big) + D \Big( \frac{1-y}{1-xy} \Big) \; = \; 0
\ee
and $d D(z) = \eta (z,1-z)$. Note, the exactness of $\eta (x,y)$ is manifest in \eqref{etaviaD},
which makes it clear that our proposed condition \eqref{quantcritK2} incorporates \eqref{qcond0}.
(The check that \eqref{quantcritK2} also incorporates \eqref{qcondQ} is similar and we leave it as an exercise to the reader.)

In our example of the $A$-polynomial for the figure-8 knot, we already claimed that the curve \eqref{curve41}
is quantizable. Indeed, the condition \eqref{qcondzz} in this example reads \cite{NeumannZagier}
\be
x \wedge y \; = \; z_1 \wedge (1 - z_1) - z_2 \wedge (1 - z_2)
\ee
where
\be
x^2 \; = \; z_1 z_2
\,, \qquad
y \; = \; \frac{z_1^2}{1-z_1} = \frac{1-z_2}{z_2^2} \,,
\ee
so that $z_1$ and $z_2$ satisfy the ``gluing condition'' $(z_1 - 1)(z_2 - 1) = z_1^2 z_2^2$.
In fact, all A-polynomials of knots have this property \cite{NeumannZagier} and, therefore, define quantizable curves 
according to our criterion (\ref{quantcritK2}).

In practice, the condition \eqref{qcondzz} is much easier to deal with
and, of course, the appearance of the dilogarithm is not an accident.
Its role in the quantization problem and the interpretation of \eqref{quantcritK2}
based on Morse theory will be discussed elsewhere \cite{inprogress}.


\subsection{The first quantum correction}
\label{sec:S1}

As we emphasized earlier, the subleading term $S_1$ contains a lot more information than meets the eye;
{\it e.g.} generically it determines much of the structure of the quantum curve, if not all of it.
Therefore, we devote an entire subsection to the discussion of $S_1$
and the first quantum correction to $\widehat{A}$ that it determines via \eqref{hier-eq2v}.

In general, the correction $S_1$ is defined as the integrated two-point function with equal arguments
$$
S_1(p) \; = \; \frac{1}{2} \int^p\int^p \omega_2(p_1,p_2) \,.
$$
The two-point function can be expressed in terms of the Bergman kernel
with a double pole removed \cite{eyn-or}
$$
\omega_2(p_1,p_2) \; = \; B(p_1, p_2) - \frac{du(p_1) du(p_2)}{\big( u(p_1) - u(p_2) \big)^2} \,.
$$
Generally, for curves of arbitrary genus, the Bergman kernel is given
by a derivative of a logarithm of the theta function of odd
characteristic $\theta_{odd}$ associated to the classical curve $\cC$ \cite{eyn-or,Marino:2006hs}
$$
B(p_1,p_2) \; = \; \partial_{p_1} \partial_{p_2} \log  \theta_{odd} \big( u(p_1) - u(p_2) \big) \,,
$$
and it has only one (second-order) pole at equal values of the arguments.
For curves of genus zero this pole is the only ingredient of
the Bergman kernel, see (\ref{Bergman}), and in that case the above
two-point function was used in (\ref{Aannulus}) to get (\ref{A1x}).

Let us discuss now how this result is modified for curves of higher
genus. For curves of genus one the Bergman kernel can be expressed
as\footnote{More generally, one can consider a generalized Bergman
kernel \cite{eyn-or}, which differes from an ordinary Bergman kernel
by a dependence on an additional parameter $\kappa$. In most
applications, including matrix models, one can set $\kappa=0$, which
leads to the ordinary Bergman kernel given above.}
\be
B(p_1,p_2) \; = \; \Big( \wp(p_1 - p_2; \tau) + \frac{\pi}{\textrm{Im} \, \tau}  \Big) dp_1 dp_2 \,.
\label{Berg-genus1}
\ee
The Weierstrass function $\wp$ has the expansion
\be
\wp(z;\tau) \; = \; \frac{1}{z^2} + \frac{g_2}{20} z^2 + \frac{g_3}{28} z^4 + \mathcal{O}(z^6) \,,
\label{WeierP}
\ee
where $\tau$ and $g_2, g_3$ denote, respectively, the modulus and the
standard invariants of an elliptic curve. Using this expansion we get
$$
\int^{p_1}\int^{p_2} \omega_2(p_1, p_2)  =
-\log\frac{u(p_1)-u(p_2)}{p_1 - p_2} + \frac{\pi}{\textrm{Im} \, \tau}
p_1 p_2 - \frac{g_2}{240} (p_1 - p_2)^4 +
\mathcal{O}\big((p_1-p_2)^6\big).
$$
In the limit $p_1 \to p_2 = p$ the first term reproduces the genus zero result \eqref{A1x},
while the other contributions in the expansion of the function $\wp(p_1-p_2;\tau)$ vanish.
In consequence, we are left with the quadratic correction to the genus zero result
\be
S_1(p) = \frac{1}{2} \int^{p_1}\int^{p_2} \omega_2(p_1, p_2) =
-\frac{1}{2} \log\frac{du}{dp} + \frac{\pi}{2 \textrm{Im} \, \tau}
p^2.  \label{S1-genus1}
\ee
As we already mentioned, for curves of higher genus the Bergman kernel also
has only one double pole at coinciding arguments. This implies that
$S_1$ for any genus will have similar structure as we found for genus one,
{\it i.e.} it will include the term (\ref{A1x}) plus some corrections.

The Bergman kernel, or the two-point function, are expressed above in
terms of uniformizing parameters $p$. Sometimes it is convenient to
express them in terms of the coordinate $u$ which enters the algebraic
equation (\ref{class-curve}) and the branch points $a_i=u(p_i^*)$
determined in (\ref{branchpoint}). For a curve of genus one there are
four branchpoints $a_1,\ldots,a_4$, and the corresponding two-point
function has been found, using matrix model techniques, in
\cite{Akemann:1996zr}. This result can also be obtained, see
\cite{Bonnet:2000dz},
using properties of elliptic functions and rewriting the Bergman
kernel given above, so that\footnote{Taking the common denominator of
the two square roots, the dependence on branch points in numerator can
be expressed in terms of symmetric functions of $a_i$, which leads to the
formula presented in \cite{DijkgraafFuji-2}.
Note that this expression, contrary to (\ref{Berg-genus1}), is manifestly holomorphic 
in the elliptic modulus $\tau$. One can adjust holomorphic dependence on $\tau$ 
by appropriate choice of the parameter $\kappa$ mentioned in the footnote above, see \cite{eyn-or}.
}
\bea
B(u_1,u_2) & = & \frac{1}{2(u_1 - u_2)^2} +
\frac{(a_3-a_1)(a_4-a_2)}{4 \sqrt{\sigma(u_1)} \sqrt{\sigma(u_2)}}
\frac{E(k)}{K(k)} + \nonumber \\
& & + \frac{1}{4(u_1-u_2)^2} \Big(\sqrt{\frac{(u_1 - a_1)(u_1-a_4)(u_2
- a_2)(u_2 - a_3)}{(u_1 - a_2)(u_1-a_3)(u_2 - a_1)(u_2 - a_4)}}  +
\nonumber \\
& & \qquad +  \sqrt{\frac{(u_1 - a_2)(u_1-a_3)(u_2 - a_1)(u_2 -
a_4)}{(u_1 - a_1)(u_1-a_4)(u_2 - a_2)(u_2 - a_3)}}  \Big),  \nonumber
\eea
where
$$
\sigma(u) = (u-a_1)(u-a_2)(u-a_3)(u-a_4)
$$
and
$$
k^2 = \frac{(a_1 - a_4)(a_2 - a_3)}{(a_1 - a_3)(a_2 - a_4)}
$$
is the modulus of the complete elliptic functions of the first and second kind, $K(k)$ and $E(k)$,
related to the parameter of the torus in (\ref{WeierP}) as $\tau=i\, K(1-k)/K(k)$.

In particular, the above expression for Bergman kernel was used in \cite{Marino:2006hs,BKMP}
to determine several terms in the $u$-expansion of the two-point function,
as well as a few lower order correlators $W^g_n$ for mirror curves of genus one,
for local $\mathbb{P}^2$ and local $\mathbb{P}^1 \times \mathbb{P}^1$.
Nonetheless, these results are not sufficient to determine corrections $\widehat{A}_1$ or $\widehat{A}_2$
to the corresponding putative quantum curves, as the hierarchy of equations (\ref{hierarchy-SA})
requires the knowledge of the exact dependence of $S_k$ on both $u$ and $v$.
We plan to elucidate this point in future work.


\section{Quantum curves and knots}
\label{sec:knots}

As we already mentioned in the introduction, in applications to knots and 3-manifolds the polynomial $A(x,y)$
is a classical topological invariant called the $A$-polynomial. (For this reason, we decided to keep the name in other examples as well
and, for balance, changed the variables to those used in the literature on matrix models and topological strings.)
In this context, the quantum operator $\widehat{A}$ is usually hard to construct (see \cite{Gelca,Garoufalidis}
for first indirect calculations and \cite{Tudor} for the most recent and systematic ones);
therefore, any insight offered by an alternative method is highly desirable.

The study of such an alternative approach was pioneered in a recent work \cite{DijkgraafFuji-2},
which focused on the computation of the perturbative partition function \eqref{Zpert}
using the topological recursion of Eynard and Orantin \cite{eyn-or}.
Starting with a rather natural\footnote{The choice of the prescription in \cite{DijkgraafFuji-2}
automatically incorporates the symmetries of the $SL(2,\C)$ character variety,
in particular, the symmetry of the $A$-polynomial under the Weyl reflection $x \mapsto x^{-1}$ and $y \mapsto y^{-1}$.}
prescription for the perturbative coefficients $S_n$ in terms of $W^g_n$,
the authors of \cite{DijkgraafFuji-2} were able to match the perturbative expansion of
the Chern-Simons partition function {\it e.g.} for the figure-8 knot complement \cite{DGLZ}
up to order $n=4$, provided certain {\it ad hoc} renormalizations are made.
It was also pointed out in \cite{DijkgraafFuji-2} that such renormalizations
are non-universal, {\it i.e.} knot-dependent. Motivated by these observations,
we start with a different prescription for the $S_n$'s described in section \ref{sec:toprecursion},
which appears to avoid the difficulties encountered in \cite{DijkgraafFuji-2} and to reproduce
the $SL(2,\C)$ Chern-Simons partition function in all examples that we checked.
In addition, we shift the focus to the $A$-polynomial itself,
and describe how its quantization \eqref{quantization}
can be achieved in the framework of the topological recursion.


\subsection{Punctured torus bundle $-L^2R^2$}

We start with a simple example of a hyperbolic 3-manifold $M$ that can be represented
as a punctured torus bundle over ${\bf S}^1$ with monodromy $\varphi = -L^2 R^2$, where
\be
L \; = \; \begin{pmatrix}
1 & 0 \cr 1 & 1
\end{pmatrix}
\qquad , \qquad
R \; = \; \begin{pmatrix}
1 & 1 \cr 0 & 1
\end{pmatrix}
\label{LRtwists}
\ee
are the standard generators of the mapping class group of a punctured torus, $\Gamma \cong PSL(2,\Z)$.
This 3-manifold has a number of nice properties. For example, it was considered in \cite{Dunfield}
as an example of a hyperbolic 3-manifold whose $SL(2,\C)$ character variety has ideal points
for which the associated roots of unity are not $\pm 1$.

For this 3-manifold $M$, the $A$-polynomial has a very simple form\footnote{In fact,
this polynomial occurs as a geometric factor in the moduli space of flat $SL(2,\C)$ connections
for infinitely many distinct incommensurable 3-manifolds \cite{Dunfield} that can be constructed
{\it e.g.} by Dehn surgery on one of the two cusps of the Neumann-Reid manifold ($=$ the unique
2-cover of $m135$ with $H_1 =\Z/2 + \Z/2 + \Z + \Z$).
Indeed, the latter is a two cusped manifold with strong geometric isolation,
which means that Dehn surgery on one cusp does not affect the shape of the other and, in particular,
does not affect the A-polynomial. As a result, all such Dehn surgeries have
the same A-polynomial $A(x,y) = 1 + ix + iy + xy$  as the manifold $m135$.}
\be
A(x,y) \; = \; 1 + ix + iy + xy \,,
\label{aLLRR}
\ee
and its zero locus, $A(x,y)=0$, defines a curve of genus zero.
According to our criterion \eqref{quantcritK2}, this curve should be quantizable.
Indeed, this can be shown either directly by verifying that all tame symbols $(x,y)_p$
are roots of unity or, alternatively \cite{RV}, by noting that the polynomial $A(x,y)$ is {\it tempered},
which means that all of its face polynomials have roots at roots of unity.
Either way, we conclude that the genus zero curve defined by the zero locus of \eqref{aLLRR}
is quantizable in the sense of section \ref{sec:ktheory}.

Therefore, we can apply the formula \eqref{A1x} from section \ref{sec:toprecursion}
to compute the one-loop correction $S_1 (u)$ or, equivalently, the torsion $T(u)$.
In fact, we can combine \eqref{Auniversal} and \eqref{A1x} to produce the following general formula
\be
\boxed{\phantom{\int^1}
\sum_{(m,n) \in \mathcal{D}}\, a_{m,n}\, c_{m,n}\, x^m y^n \; = \;
\frac{1}{2} \Big( \frac{du}{dp} \Big)^{-2} \, \left(
\frac{d^2 u}{dp^2} \, \partial_v
- \frac{du}{dp} \, \frac{dv}{dp} \, \partial_v^2 \right) \; A
\phantom{\int^1}}
\label{Agenus0}
\ee
that allows to determine the exponents $c_{m,n}$ of the $q$-deformation \eqref{Aqnice}
directly from the data of the classical $A$-polynomial $A = \sum a_{m,n} x^m y^n$
and a parametrization \eqref{param}.

In our present example, we can choose the following parametrization:
\be
x (p) =  - \frac{1 + ip}{i + p}
\qquad , \qquad
y(p) = p \,,
\ee
suggested by the form of \eqref{aLLRR}. Substituting it into \eqref{Agenus0} uniquely determines
the values of the $q$-exponents $c_{m,n}$ and, therefore, the quantum operator \eqref{Aqnice}:
\be
\widehat{A} \; = \; 1 + i q^{1/2} \hat x + i q^{-1/2} \hat y + q \hat x \hat y \,.
\label{ahatLLRR}
\ee
In order to fully appreciate how simple this derivation of $\widehat{A}$ is
(compared to the existent methods and to the full-fledged topological recursion)
it is instructive to follow through the steps of sections \ref{sec:toprecursion} and \ref{sec:hierarchy}
that, eventually, lead to the same result \eqref{ahatLLRR}.

First, one needs to go through all the steps of the topological recursion.
Relegating most of the details to section \ref{sec-conifold},
where \eqref{aLLRR} will be embedded in a larger class of similar examples
(and dealing with various singular limits as presented in section \ref{sec-tetrahedron}),
we summarize here only the output of \eqref{conjduality}:
\bea
S_0' & = & \log \frac{x-i}{ix-1} \,, \nonumber \\
S_1' & = & \frac{i-x}{2x+2i} \,, \nonumber \\
S_2' & = & \frac{x (5i -12x - 5i x^2)}{12(1 + x^2)^2} \,, \nonumber \\
& \vdots & \nonumber
\eea
which should be used as an input for \eqref{hierarchy-SA}. Indeed, from the first
few equations in \eqref{Aclass}-\eqref{hier-SA-Cstar} one finds the perturbative expansion \eqref{Ahatpert}
of the quantum operator $\widehat{A}$:
\bea
\widehat{A}_1 & = & \frac{1}{2} \big( i \hat x - i \hat y + 2 \hat x \hat y \big) \,, \nonumber \\
\widehat{A}_2 & = & \frac{1}{8} \big( i \hat x + i \hat y + 4 \hat x \hat y \big) \,, \nonumber \\
\widehat{A}_3 & = & \frac{1}{48} \big( i \hat x - i \hat y + 8 \hat x \hat y \big) \,, \nonumber \\
& \vdots & \nonumber
\eea
It does not take long to realize that the perturbative terms $\widehat{A}_n$ come from
the $\hbar$-expansion of the ``quantum polynomial'' \eqref{ahatLLRR} with $q = e^{\hbar}$.
Pursuing the topological recursion further, one can verify this to arbitrary order in
the perturbative $\hbar$-expansion, thus, justifying that $\widehat{A}$
can be written in a nice compact form \eqref{Aqnice}.

Hence, our present example provides a good illustration of how all
these steps can be streamlined in a simple computational technique \eqref{uvgivesA} which,
for curves of genus zero, can be summarized in a single general formula \eqref{Agenus0}.



\subsection{Figure-8 knot}     \label{ssec-figure8}

The lesson in our previous example extends to more interesting knots and 3-manifolds,
sometimes in a rather trivial and straightforward manner and, in some cases, with small new twists.
The main conceptual point is always the same, though:
at least in all examples that "come
from geometry," {\em the full quantum curve $\widehat{A}$ is completely
determined by the first few terms in the $\hbar$-expansion,
which can be easily obtained using the tools of the topological recursion.}

For example, let us consider the figure-8 knot complement, $M = {\bf S}^3 \setminus K$,
for which the story is a little less trivial. The figure-8 knot is shown in figure \ref{fig:knot41}.
Much like our first example in this section, $M$ is a hyperbolic 3-manifold that also can be
represented as a punctured torus bundle with the monodromy
\be
\varphi = RL =
\begin{pmatrix}
2 & 1 \cr 1 & 1
\end{pmatrix} \,,
\label{figeightd}  \nonumber
\ee
where $L$ and $R$ are defined in \eqref{LRtwists}.
Even though the classical curve \eqref{curve41} for the figure-8 knot
is hyper-elliptic, one can still easily find the torsion $T(u)$ needed for \eqref{Auniversal}.
In fact, for curves associated\footnote{{\it i.e.} defined
by the zero locus of the $A$-polynomial} with knots and 3-manifolds the torsion $T(u)$ is exactly what low-dimensional topologists
call the Ray-Singer (or Reidemeister) torsion of a 3-manifold $M$.
To be more precise, the function $T(u)$ is the torsion of $M$ twisted by a flat $SL(2,\C)$ bundle $E_{\rho} \to M$
determined by the representation $\rho : \pi_1 (M) \to SL(2,\C)$ or, at a practical level,
by the point $\rho = (x,y)$ on the classical curve $\cC$.

\FIGURE{\;\;\includegraphics[width=1.4in]{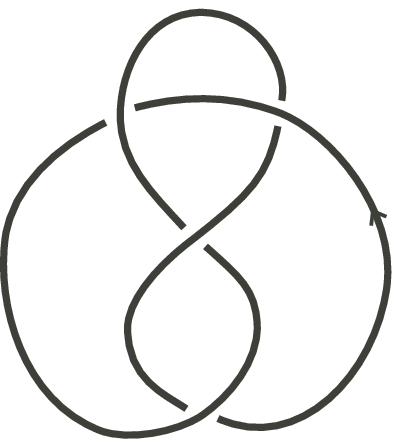}
\caption{Figure-8 knot. \\}\label{fig:knot41}
}

In particular, $T(u)$ is a topological invariant of $M = {\bf S}^3 \setminus K$ and,
therefore, can be computed by the standard tools.
For instance, when $\rho$ is Abelian, the torsion $T(u)$ is related to
the Alexander-Conway polynomial $\nabla (K;z)$ \cite{Milnor,Turaev}:
\be
\sqrt{T} \; = \; \frac{\nabla (K;x - x^{-1})}{x-x^{-1}}
\ee
that, for every knot $K$, can be computed by recursively applying a simple skein relation\footnote{For example,
$\nabla_{{\bf 3}_1} (z) = 1+z^2$ for the trefoil knot and $\nabla_{{\bf 4}_1} (z) = 1 - z^2$ for the figure-8 knot.
Note, that our definition of $T(u)$ is actually the inverse of the Ray-Singer torsion, as defined in the mathematical literature.
This unconventional choice turns out to be convenient in other applications, beyond knots and 3-manifolds.}
\be
\nabla (\overcrossing) - \nabla (\undercrossing) = z \, \nabla (\smoothing)\,,
\label{skein}
\ee
and the normalization $\nabla (\unknot) = 1$.
Similarly, when $\rho$ is non-Abelian (and irreducible) the torsion looks like
\be
T (x) \; = \; \sqrt{ \Delta (x) } \,,
\label{twistedtorsion}
\ee
where $\Delta (x)$ is the Alexander polynomial of $M$ twisted by the flat $SL(2,\C)$ bundle $E_{\rho}$, cf. \cite{surveyAlexander}. For example, for the figure-8 knot that we are interested in here, it has the form \cite{Porti,GMtorsion}:
\be
\Delta_{{\bf 4}_1} (x) \; = \; - x^{-4} + 2 x^{-2} + 1 + 2 x^2 - x^4 \,.
\label{t41}
\ee

Now we are ready to plug this data into our universal formula \eqref{Auniversal} and compute
the quantum operator $\widehat A$ or, at least, its first-order approximation.
The computation is fairly straightforward; indeed, from \eqref{twistedtorsion} and \eqref{t41} we find
\be
\frac{\partial_u T}{T} \; = \;
\frac{2(-1+x^2-x^6+x^8)}{1-2x^2-x^4-2x^6+x^8}
\ee
and, by solving \eqref{curve41} we get
$y^{(\alpha)} (x) = \frac{1 - x^2 - 2 x^4 - x^6 + x^8}{2x^4} \pm \frac{1-x^4}{2x^2} \sqrt{-\Delta (x)}$
which immediately gives the second part of the input data for \eqref{Auniversal}, namely
\be
\frac{\partial_u A}{\partial_v A} = - \frac{dv}{du}
= \frac{2(2x^{-2}-1+2x^2)}{\sqrt{-\Delta (x)}} \,.
\ee
Then, once we plug these ingredients into \eqref{Auniversal} we come to our first surprise:
we find that there is no way to satisfy \eqref{Auniversal} with constant real numbers $c_{m,n}$
if for $\mathcal{D}$ we simply take the Newton polygon of the classical curve \eqref{curve41}.
In other words, the figure-8 knot is a good illustration of the following phenomenon (that rarely
happens in simple examples, but seems to be fairly generic in more complicated ones):
one may need to enlarge the domain $\mathcal{D}$ in order to solve \eqref{Auniversal}.
For the figure-8 knot, the minimal choice is
\be
A(x,y) \; = \; (1 - x^4) x^4 - (1 - x^4) (1 - x^2 - 2 x^4 - x^6 + x^8) y + (1 - x^4) x^4 y^2
\ee
and differs from \eqref{curve41} by an extra factor $1-x^4$.
Now, with this $A(x,y)$, the formula \eqref{Aleading} produces the set of coefficients $c_{m,n}$ or,
equivalently, their ``generating function''
\be
\widehat{A}_1 \; = \;
(3 - 9 \hat x^4) \hat x^4
- ( - 2 \hat x^2 - 12 \hat x^4 + 24 \hat x^8 + 10 \hat x^{10} - 12 \hat x^{12}) \hat y
+ (5 - 7 \hat x^4) \hat x^4 \hat y^2 \,,
\ee
which almost uniquely determines the full quantum $A$-polynomial for the figure-8 knot in Table \ref{table}:
$$
\widehat{A} \; = \;
q^{3} (1-q^6 \hat{x}^4) \hat{x}^4
- (1 - q^4 \hat{x}^4) (1 - q^2 \hat{x}^2 - (q^2 + q^6) \hat{x}^4 - q^6 \hat{x}^6 + q^8 \hat{x}^8) \hat{y}
+ q^{5} (1 - q^2 \hat{x}^4) \hat{x}^4 \hat{y}^2 \,.
$$
Indeed, if one knows that $\widehat{A}$ is in the general form \eqref{Aqnice},
then the above expression for $\widehat{A}_1$ determines almost all of the coefficients in $\widehat{A}$,
except for the factor $q^2+q^6$ which is easily fixed by going to the next order in the recursion.


\subsection{Torus knots and generalizations}
\label{sec:torus}

For a $(m,n)$ torus knot, the classical curve \eqref{class-curve} is defined by a very simple polynomial \cite{CCGL}:
\be
A (x,y) \; = \; y - x^{mn} \,.
\label{Atorus}
\ee
In fact, this curve is a little ``too simple'' to be an interesting example
for quantization since it has only two monomial terms, whose relative coefficient
in the quantum version
\be
\widehat{A} (\hat x,\hat y) \; = \; \hat y - q^c \hat x^{mn}
\label{Atorusquantum}
\ee
can be made arbitrary by a suitable canonical transformation, as discussed in section \ref{sec:choices}.
(Indeed, one can attain arbitrary values of~$c$ even with the simple shift transformation \eqref{vshift}.)
Another drawback of \eqref{Atorus} is that, for general $m$ and $n$, it describes a singular curve.

Both of these problems can be rectified by passing to a more general class of examples,
\be
A (x,y) \; = \; y + P(x) \,,
\label{Ator}
\ee
where $P(x)$ can be either a polynomial or, more generally, an arbitrary function of $x$.
Then, the $A$-polynomial \eqref{Atorus} of $(m,n)$ torus knots (and its quantization \eqref{Atorusquantum})
can be recovered as a limiting case of this larger family, $P(x) \to - x^{mn}$.
Another important advantage of choosing generic $P(x)$ is that we can use \eqref{Agenus0} to find $\widehat{A}$.

In practice, in order to implement the algorithm summarized in \eqref{uvgivesA} and \eqref{Agenus0},
it is convenient to exchange the role of $x$ and $y$. Hence, we will work with the ``mirror'' version of~\eqref{Ator}:
\be
A (x,y) \; = \; x + P(y) \,,
\label{Alinear}
\ee
where $P(y)$ can be an arbitrary function of $y$.
In general, the curve defined by the zero locus of this function is a multiple cover of the $x$-plane.
It admits different parametrizations which, therefore, lead to different expressions for $\widehat{A}$
(related by canonical transformations discussed in section \ref{sec:choices}).
However, one can always make a natural choice of parametrization with
\be
\left\{\begin{array}{l} x(p) = -P(p) \\
y(p) =  p \end{array}\right.   \label{xPy-param}
\ee
Substituting this into \eqref{Auniversal} (or, equivalently, into \eqref{Agenus0}) we find
\be
\sum_{(m,n) \in \mathcal{D}}\, a_{m,n}\, c_{m,n}\, x^m y^n \; = \;
\frac{x}{2} - \frac{y}{2} \frac{dP(y)}{dy}
\ee
which, for generic $P(y)$, immediately determines the quantization of \eqref{Alinear}:
\be
\boxed{\phantom{\int^1}
\widehat{A} \; = \; q^{1/2} \hat x + P \big( q^{-1/2} \hat y \big) \,.
\phantom{\int^1}}
\label{xPy-Ahat}
\ee
Notice, in spite of the suggestive notation, $P(y)$ does not need to be a polynomial in this class of examples.
For instance, choosing $P(y)$ to be a rational function,
\be
P(y) \; = \; \frac{1 + iy}{i + y}
\ee
{}from \eqref{xPy-Ahat} we find the quantum curve,
\be
q^{1/2} \hat x + \frac{ q^{1/2} + i \hat y}{i q^{1/2} + \hat y} \; \simeq \; 0 \,,
\ee
which, after multiplying by $i q^{1/2} + \hat y$ on the left
and using the commutation relation $\hat y \hat x = q \hat x \hat y$,
agrees with the earlier result \eqref{ahatLLRR}.


\section{Examples with $\widehat{A} = A_{{\rm classical}}$}

In certain examples, it turns out that the quantum curve can be obtained from the classical one
simply by replacing $u$ and $v$ by $\hat{u}$ and $\hat{v}$ with no additional $\hbar$ corrections
(and with our standard ordering conventions, {\it cf.} section \ref{sec-intro}).
There are examples of such special curves in $\C\times \C$ as well as in $\C^*\times\C^*$;
{\it e.g.} from \eqref{xPy-Ahat} it is easy to see that $A(x,y) = x + 1/y$ is one example.
In this section, for balance, we consider curves with this property defined by a polynomial
equation $A(u,v)=0$ in $\C\times \C$.
In particular, we discuss in detail a family of examples related to the Airy function,\footnote{
In this model, computation of $W^g_n$ and their generating functions are also presented in \cite{eyn-or}.}
in order to explain how our formalism works for curves embedded in $\C\times \C$.

The Airy function (and its cousins) can be defined by a contour integral,
\be
Z_{{\rm Ai}} (u) \; = \; \int_{\gamma} \frac{dz}{2\pi i} \, e^{-\frac{1}{\hbar} S(z)} \,, \qquad S(z) = -uz + \frac{z^3}{3}   \label{Ai-contour}
\ee
over a contour $\gamma$ that connects two asymptotic regions in the complex $z$-plane
where the ``action'' $S$ behaves as ${\rm Re}\, S (z) \to + \infty$.
For such a contour $\gamma$, we have the following Ward identity:
$$
0 = \frac{1}{2\pi i} \int_{\gamma} d \left[ e^{-\frac{1}{\hbar} S(z)} \right]
= \frac{1}{\hbar} \int_{\gamma} \frac{dz}{2\pi i} \left( u - z^2 \right)  e^{-\frac{1}{\hbar} S(z)}
$$
which we can write in the form of the differential equation
\be
\left( \hat{v}^{\, 2} - u \right) Z_{{\rm Ai}} (u) \; = \; 0              \label{Airy-diffeq}
\ee
where we used the definition of $Z_{{\rm Ai}} (x)$ and
\be
\hat{v}^{\, 2} Z_{{\rm Ai}} (x) = \left( \hbar \partial_u \right)^2 \int_{\gamma} \frac{dz}{2\pi i} \, e^{- \frac{1}{\hbar} S(z)}
= \int_{\gamma} \frac{dz}{2\pi i} \, z^2 \, e^{-\frac{1}{\hbar} S(z)} \,.
\ee

This simple, yet instructive, example is a prototype for a large class of models
where quantum curves are identical to the classical ones, {\it i.e.} $\hat A = A (u,v)$.
Indeed, let us consider a contour integral,
$$
Z (u) \; = \; \int_{\gamma} \frac{dz}{2\pi i} \, e^{-\frac{1}{\hbar} S(z)} \,, \qquad S(z) = -uz + P(z)
$$
where $\gamma$ is a suitable contour in the complex $z$-plane, and $P(z)$ is a Laurent polynomial.
Then, following the same arguments as in the example of the Airy function,
we obtain the following Ward identity
$$
\int_{\gamma} \frac{dz}{2\pi i} \left( u - P'(z) \right)  e^{-\frac{1}{\hbar} S(z)} = 0
$$
which translates into a differential equation $\hat A Z(u) = 0$ with
\be
\hat A = P' (\hat v) - \hat u \,.
\label{apu}
\ee
The special choice of $P'(z) = z^p$ gives rise to $(p,1)$ minimal model coupled to gravity.
In this case, the corresponding partition function has an interpretation of the amplitude
of the FZZT brane \cite{Maldacena:2004sn}, and in the dual matrix model this partition function
is indeed computed as the expectation value of the determinant (\ref{det-uM}).
Recall, that a double scaling limit of hermitian matrix models with polynomial potentials
describes $(p,q)$ minimal models coupled to gravity, characterized by singular spectral curves \cite{DFGZJ}:
\be
A(u,v) = v^p - u^q = 0 \,.
\label{minimalpq}
\ee
In the simpler case of $q=1$ discussed here the classical Riemann surface $P'(v)-u=0$,
given by the $\hbar\to 0$ limit of the quantum curve \eqref{apu},
represents the semi-classical target space of the minimal string theory.
Below we discuss in detail how the above $\hat A$ arises from our formalism in the Airy case, $p=2$.

\subsection{Quantum Airy curve}

For a minimal model with $(p,q)=(2,1)$ the classical curve \eqref{minimalpq} looks like
\be
A(u,v) = v^2 - u = 0 \,.
\label{AiryA}
\ee
It has two branches labeled by $\alpha=\pm$,
\be
v = S'_0 = \pm\sqrt{u}=v^{(\pm)} \,,
\label{Airy-branches}
\ee
and exchanged by the Galois transformation\footnote{By definition,
the action of the Galois group preserves the form of the curve (\ref{AiryA}).}
$$
v \to -v \,.       
$$

This model provides an excellent example for illustrating how
the hierarchy of differential equations (section \ref{sec:hierarchy})
and the topological recursion (section \ref{sec:toprecursion}) work.
Because we already know the form of the quantum curve in this example,
we start by deriving the $\hbar$ expansion of the Airy function using the hierarchy (\ref{hierarchy-SA}).
Then, we will show that this expansion is indeed reproduced by the topological recursion.
In examples considered later we will also illustrate the reverse process: from the knowledge of $S_k$
(computed from the topological recursion) we will determine the form of the quantum curve.

In our calculations, we will use global coordinates, such as $v$ or $p$, and avoid
using the coordinate $u$ (that involves a choice of branch of the square root) except for writing the final result.
In particular, from the equation of the Airy curve (\ref{AiryA}) we find the relation
\be
v'=\frac{dv}{du} = -\frac{\partial_u A(u,v)}{\partial_v A(u,v)} = \frac{1}{2v}
\label{Airy-v-prim}
\ee
that will be useful below.

\subsubsection{Differential hierarchy}

First, we solve the hierarchy of equations that follow from the quantum curve (\ref{Airy-diffeq}):
\be
\widehat{A} Z_{{\rm Ai}}= \big(\hbar^2\partial_u^2 - u\big)Z_{{\rm Ai}} = 0\, .    \label{Airy}
\ee
To solve this equation in variable $u$, already in the first step one would have to make a choice of the branch (\ref{Airy-branches}).
This would influence then all higher order equations in the differential hierarchy,
and eventually lead to two well-known variants of the Airy function.
Instead, we express the coefficients $S_k$ in a universal way in terms of $v$,
so that a particular solution in terms of $u$ can be obtained by evaluating $v$ in the final expression
on either branch~\eqref{Airy-branches}.

The first equation in the differential hierarchy is already given in (\ref{Airy-branches}), {\it i.e.} $v=S'_0$.
The second equation (\ref{hier-eq2v}) takes form
$$
S'_{1} \partial_v A(u,v) + \frac{1}{2} S''_{0} \partial_v^2 A(u,v) = 0 \,,
$$
and implies
\be
S'_{1} = -\frac{v'}{2v} = -\frac{1}{4v^2}.            \label{S1-mm21} \nonumber
\ee
Solving further equations (\ref{hierarchy-SA}) we find
\be
S'_2 = \frac{-1-8v v'}{32 v^5} = -\frac{5}{32v^5}
\,, \qquad
S'_3 = -\frac{5(1 + 10v v')}{128v^8} = -\frac{15}{64v^8}
\,, \qquad
S'_4 = -\frac{1105}{2048v^{11}} \,.
\nonumber
\ee
We can integrate these results taking advantage of (\ref{Airy-v-prim}) to find
\be
S_k = \int \frac{S'_k}{v'}dv \,.     \label{intSprim}
\ee
In particular, the first few terms look like
\be
S_0 = \frac{2}{3}v^3 \,,\quad S_1 = -\frac{1}{2} \log \,v
\,,\quad  S_2 = \frac{5}{48v^3} \,, \quad
S_3 = \frac{5}{64v^6} \,, \quad
S_4 = \frac{1105}{9216v^9} \,.
\label{Airy-Sk-diffhier}
\ee
Finally, using (\ref{Airy-branches}) we can evaluate these expressions on either of the two branches $v^{(\pm)}=\pm\sqrt{u}$
to find two asymptotic expansions of the Airy function~\eqref{Ai-contour} (often denoted Bi and Ai),
\be
Z^{(\pm)}_{\textrm{Ai}}(u)  = \frac{1}{u^{1/4}} \exp\Big(
  \pm\frac{ 2 u^{3/2}}{3 \hbar} \pm \frac{5 \hbar}{48 u^{3/2}} + \frac{5 \hbar^2}{64 u^3} \pm
    \frac{1105 \hbar^3}{9216 u^{9/2}} + \ldots \Big) \,,
    \label{Z-Airy}
\ee
which indeed satisfy the second order equation (\ref{Airy}).

\subsubsection{Topological recursion}

Now we reconsider the Airy curve from the topological recursion viewpoint. The classical curve can be parametrized as
\be
\left\{\begin{array}{l} u(p) = p^2  \\
v(p) =   p \end{array}\right.   \nonumber   
\ee
The conjugate point is simply $\bar{p} = -p$, and there is one branch-point at $p=0$.
All ingredients of the recursion can be found in the exact form, in particular
the anti-derivative and the recursion kernel take the following form
$$
S_0 (p) \; = \; \int^p \phi = \frac{2}{3} p^3 \,,
\qquad \qquad   K(q,p) = \frac{1}{4q(p^2 - q^2)} \,.
$$
The annulus amplitude gives
$$
S_1 = -\frac{1}{2}\log\frac{du}{dp} = -\frac{1}{2} \log(2v) \,,
$$
which correctly reproduces $S_1$ found in (\ref{Airy-Sk-diffhier}) (up to an irrelevant constant).

Now we apply the topological recursion to find the higher order terms $S_k$ with $k \ge 2$.
These terms are computed as functions on the curve, {\it i.e.} as functions of the parameter $p$,
and can be expressed as rational functions of $u$ and $v$.
In particular we find
$$
W^1_1(p) = -\frac{1}{16p^4}, \qquad \qquad W^0_3(p_1,p_2,p_3) = -\frac{1}{2p_1^2 p_2^2 p_3^2} \,,
$$
which implies
$$
S_2 = \int^p W^1_1(p) dp + \frac{1}{6} \iiint^p W^0_3(p_1,p_2,p_3) dp_1 dp_2 dp_3 = \frac{5}{48v^3} \,.
$$
In higher orders, we get
$$
S_3 = \frac{5}{64v^6} \,, \qquad \qquad  S_4 = \frac{1105}{9216v^9} \,.
$$
These results agree with the expansion (\ref{Airy-Sk-diffhier}) obtained from the differential hierarchy.
It is clear that, had we not known the form of the quantum curve to start with,
we could compute the coefficients $S_k$ using the topological recursion and then
apply the hierarchy of differential equations (\ref{hierarchy-SA}).
This would reveal that all quantum corrections $\hat A_k$ vanish,
and the quantum curve indeed takes the form (\ref{Airy}) and coincides with the classical curve.

Let us also illustrate the factorization of the quantum curve (\ref{Ahat-factor}) to the leading order in $\hbar$.
In the polarization where $p$ is the ``coordiante,'' the curve (\ref{Airy}) takes the form
$$
\widehat{A} = \big( \hbar\partial_p - 2p^2  \big)\big(  \hbar  \partial_p + 2p^2  \big) + \mathcal{O}(\hbar) \,.
$$
Then, to the leading order, the two branches of the partition function are annihilated by the operators $(\hbar \partial_p \mp 2^2p )$
and the solutions to these equations represent the two variants of the Airy function (\ref{Z-Airy}):
$$
Z = e^{\pm \frac{2p^3}{3\hbar}}\big( 1 + \mathcal{O}(\hbar) \big) = e^{\pm \frac{2u^{3/2}}{3\hbar}}\big( 1 + \mathcal{O}(\hbar) \big) \,.
$$


\section{$c=1$ model}

The aim of this section is to analyze the so-called $c=1$ model.
As in the previous section, however, it is instructive to start with a more a general class of models associated with the contour integral
$$
Z (u) \; = \; \int_{\gamma} \frac{dz}{2\pi i} \, z^{\frac{t}{\hbar}} \, e^{-\frac{1}{\hbar} S(z)} \,,
\qquad S(z) = -uz + \frac{z^{n+1}}{n+1}  \,.
$$
This integral satisfies the following Ward identity
$$
\int_{\gamma} \frac{dz}{2\pi i} \left( \frac{t}{z} + u - z^n \right) z^{\frac{t}{\hbar}}  e^{-\frac{1}{\hbar} S(z)} = 0
$$
that leads to the quantum curve
$$
\hat A = t + \hat v \left( \hat u - \hat v^{\, n} \right)  \,.
$$
In the special case $n=1$ this reproduces the quantum curve of the $c=1$ model:
$$
\hat A = t + \hat v \hat u
$$
where we used the freedom of shifting $u$ by an arbitraty function of $v$
to implement a change of polarization $\hat u \to \hat u + \hat v$, {\it cf.} section \ref{sec:choices}.
(Note that this shift does not affect the commutation relations of $\hat u$ and~$\hat v$.)
Another convenient choice of polarization is implemented by a canonical transformation
$$
\hat u \to \frac{1}{\sqrt{2}} \left( \hat u - \hat v \right)
\,, \qquad
\hat v \to \frac{1}{\sqrt{2}} \left( \hat u + \hat v \right)
$$
and leads to a perhaps more familiar representation of the quantum curve for the $c=1$ model:
\be
\widehat{A} = \left( \hat u + \hat v \right) \left( \hat u - \hat v \right) + 2t
= \hat u^{\, 2} - \hat v^{\, 2} + 2 t + \hbar \,.     \label{c1-Ahat}
\ee

In what follows we consider this last form of the quantum curve.
Note, in this case the underlying classical curve is embedded in $\C\times\C$ by the equation
\be
A(u,v) = u^2 - v^2 + 2t =0 \,,   \label{c1-model}
\ee
and has two branches $v^{(\alpha)}$ labeled by $\alpha=\pm$,
\be
v^{(\pm)}(u) = \pm \sqrt{u^2 + 2t} \,.   \label{c1model-vu}
\ee
These branches are mapped to each other by a Galois transformation
$$
v\to -v \,,
$$
that does not change the form of the curve (\ref{c1-model}). We also note that
\be
v' = \frac{dv}{du} = -\frac{\partial_u A(u,v)}{\partial_v A(u,v)} = \frac{u}{v} \,.    \label{c1-v-prim}
\ee

The solution of the $c=1$ model is well known. In particular, the associated closed string free energies, for $g\geq 2$, are given by
\be
F_g = \frac{B_{2g}}{2g(2g-2)} \frac{1}{t^{2g-2}} \,.   \label{Fgc1}
\ee
Below we reexamine this model in the new formalism, in particular from the viewpoint of open (rather than closed) string invariants.
Since the quantum curve (\ref{c1-Ahat}) has only the first order quantum correction $\hat A_1 = 1$,
we start by verifying that it is indeed correctly reproduced by the annulus amplitude (\ref{A1x}) in our formalism.
Then, we follow the strategy employed in the previous section and show that higher order amplitudes $S_k$,
determined by the quantum curve equation, agree with those given by the topological recursion.
Equivalently, this guarantees that, had we computed $S_k$ first by applying the topological recursion
to the classical curve (\ref{c1-model}) and then determined the quantum curve using the hierarchy (\ref{hierarchy-SA}),
we would indeed find $\widehat{A}$ given in (\ref{c1-Ahat}).


\subsection{Differential hierarchy}

The differential hierarchy (\ref{hierarchy-SA}) starts with the equation which, as usual,
specifies the disk amplitude; integrating (\ref{c1model-vu}) we find that it takes the form
\be
S_0 = \pm \Big( \frac{1}{2}u\sqrt{u^2 + 2t}  +  t \log(u+\sqrt{u^2+ 2t}) \Big)  \,.  \label{S0-c1}
\ee
The second equation in the differential hierarchy (\ref{hier-eq2v})
implies that
\be
S'_1 = \frac{A_1 - v'}{2 v} = \frac{A_1 v - u}{2 v^2} \,,       \label{S1c1}  
\ee
with $A_1=1$. The first (and the only) quantum correction $A_1=1$ follows directly from (\ref{c1-Ahat})
as well as from the the annulus amplitude which we compute below in (\ref{S1c1TR}).

To find the higher order amplitudes $S_k$ we take advantage of the fact that
all higher order corrections to the quantum curve (\ref{c1-Ahat}) vanish.
Therefore, using the fact that the first correction $A_1=1$ is annihilated by all $\D_{r>0}$,
all higher order equations in the hierarchy (\ref{hierarchy-SA}) take a simple form $\D_n A=0$.
Moreover, noting that the classical curve is quadratic in $v$, the hierarchy of differential equations reduces to
\bea
0 & = & S'_2 \partial_v A + \frac{1}{2} \big((S'_1)^2 + S''_1\big)  \partial_v^2 A \,,  \nonumber \\
0 & = & S'_3 \partial_v A + \Big(\frac{S''_2}{2} + S'_1 S'_2  \Big) \partial_v^2 A \,,   \nonumber \\
0 & = & S'_4 \partial_v A + \frac{1}{2}\Big((S'_2)^2 + S''_3 + 2S'_1 S'_3  \Big)    \partial_v^2 A \,,   \nonumber \\
& \vdots & \nonumber
\eea
and solving these equations we get
\bea
S'_1 & = & \frac{v - u}{2 v^2} \,, \nonumber \\
S'_2 & = & \frac{-5u^2 + 4uv + v^2}{8 v^5} \,, \label{S2c1}    \\
S'_3 & = & -\frac{(u-v)(3u-v)(2u+v)}{16 v^8} \,,   \nonumber \\
S'_4 & = & -\frac{(u - v) (1105 u^3 + 145 u^2 v - 389 u v^2 - 21 v^3)}{128 v^{11}} \,.    \nonumber
\eea
We stress that given here are global solutions; in order to restrict to a particular
branch one needs to substitute $v=v^{(\pm)}$ using (\ref{c1model-vu}).
Making such a choice of branch and expanding in $u$ we find
\bea
S'_{1,\pm} & = & \pm \frac{1}{2\sqrt{2t}} - \frac{u}{4t} \mp \frac{u^2}{4(2t)^{3/2}} + \frac{u^3}{8t^2}  \pm\frac{3u^4}{16(2t)^{5/2}} - \frac{u^5}{16t^3} + \ldots \nonumber\\
S'_{2,\pm} & = & \pm \frac{1}{8 (2t)^{3/2}} + \frac{u}{8 t^2} \mp \frac{13 u^2}{16 (2t)^{5/2}} - \frac{u^3}{8t^3}
\pm \frac{115 u^4}{64 (2t)^{7/2}} + \frac{3 u^5}{32 t^4} +\ldots  \nonumber \\
S'_{3,\pm} & = &   \mp \frac{5}{16 (2t)^{5/2}} + \frac{5 u}{64 t^3} \pm \frac{75 u^2}{32 (2t)^{7/2}} - \frac{15 u^3}{64 t^4} \mp \frac{875 u^4}{128 (2t)^{9/2}} + \frac{45 u^5}{128 t^5}   + \ldots \nonumber \\
S'_{4,\pm} & = &
\mp \frac{21}{128 (2t)^{7/2}} - \frac{23 u}{128 t^4} \pm\frac{1215 u^2}{256 (2t)^{9/2}}
+ \frac{19 u^3}{32t^5} \mp \frac{29387 u^4}{1024 (2t)^{11/2}} - \frac{265 u^5}{256 t^6} + \ldots  \nonumber
\eea
Integrating these results term by term gives the $u$ expansion of $S_k$.
One can also find the global representation of $S_k$ in terms of $u$ and $v$
using the integral (\ref{intSprim}) and the result (\ref{c1-v-prim});
we determine such a global representation below.


\subsection{Topological recursion}

Now we show how the above results can be reproduced using the topological recursion. The curve (\ref{c1-model}) can be parametrized as
\be
\left\{\begin{array}{l} u(p) = 2 p t - \frac{1}{4p} \\
v(p) =   2p t + \frac{1}{4p} \end{array}\right.
\ee
Note, this implies that a local parameter $p$ can be expressed as
$$
p = \frac{u+v}{4t}  \,.
$$

Having fixed the parametrization, we can compute the annulus amplitude (\ref{A1x}) and find that its derivative in this case is
\be
S'_1 = \frac{v-u}{2v^2} \,.  \label{S1c1TR}
\ee
Comparing this with (\ref{S1c1}) we confirm that the first quantum correction to the $A$-polynomial indeed reads
$$
A_1 =  1 \,,
$$
in complete agreement with (\ref{c1-Ahat}).

The other ingredients of the topological recursion are as follows. There are two branch points $du(p_*)=0$ at
\be
p_* = \pm \frac{i}{2\sqrt{2t}} \,.
\ee
Conveniently, there is a global expression for the conjugate point
\be
\overline{p} = - \frac{1}{8tp} \,,
\ee
and the recursion kernel reads
\be
K(q,z) = \frac{4q^3}{(1 - 8 q^2 t) (q - z) (1 - 8 q t z)} \,.
\ee
The correlators contributing to (\ref{Agn-p}) take form
\bea
W^1_1(p) & = &  \frac{64 p^3 t}{(1+ 8 p^2 t)^4} \,,  \nonumber \\
W^2_1(p) & = &  \frac{86016 t (p^7-24 p^9 t+64 p^{11} t^2)}{(1+8 p^2 t)^{10}} \,,   \nonumber\\
W^3_1(p) & = &  \frac{2883584 p^{11} t (135 - 8784 p^2 t + 133376 p^4 t^2 - 562176 p^6 t^3 +
   552960 p^8 t^4)}{(1+8 p^2 t)^{16}} \,,   \nonumber
\eea
and so on.
Hence, using (\ref{Agn-p}) we get the global representation
$$
S_2 = \frac{2t (2t - 9 (u + v)^2)}{6 (2t + (u + v)^2)^3} \,, \qquad \quad
S_3 = \frac{20 t (u + v)^4 (2t - (u + v)^2)}{(2t + (u + v)^2)^6} \,,
$$
and derivatives of these functions with respect to $u$ indeed agree with our earlier results~\eqref{S2c1}. 
Therefore, the results of the topological recursion are in excellent agreement with~\eqref{c1-Ahat}.
Again, had we not known the quantum curve to start with, we could reverse the order of the computation
and from the knowledge of the coefficients $S_k$ determine
\be
\widehat{A} = u^2 - (\hbar \partial_u)^2 +2t + \hbar \,.
\ee
Finally, we illustrate the factorization property (\ref{Ahat-factor}) of the quantum curve in $p$-polarization.
In this polarization, the curve (\ref{c1-Ahat}) gives rise to first order differential operators
$\big( \hbar \partial_p \mp \frac{(1+8tp^2)^2}{16p^3} \big)$
which (to the leading order in $\hbar$) annihilate the two branches of the partition function:
$$
Z^{(\alpha)} = e^{\pm\frac{1}{\hbar}\big(-\frac{1}{32p^2} + 2t^2 p^2 + t \log p \big) }   \Big( 1 + \mathcal{O}(\hbar) \Big) \,.
$$
After substituting $p=(u+v)/4t$ and $v$ given by (\ref{c1model-vu})
we indeed reproduce the leading behavior (\ref{S0-c1}).

Let us also mention that from $W^2_1$ and $W^3_1$ computed here one can determine
the ``closed string'' free energies (\ref{Fg}). This computation reveals that
$$
F_2 = -\frac{1}{240t^2} \,, \quad \qquad F_3 = \frac{1}{1008t^4}
$$
in excellent agreement with the expected result (\ref{Fgc1}), thereby,
providing yet another nice check of the topological recursion formalism.


\section{Tetrahedron or framed $\C^3$}
\label{sec-tetrahedron}

In this section we consider quantization of a classical curve that
plays a very important role simultaneously in two different areas:
in low-dimensional topology and in topological string theory.

One of the problems in low-dimensional topology is to associate quantum group
invariants to 3-manifolds (possibly with boundary).
Topological string theory, on the other hand, computes various enumerative
invariants of Calabi-Yau 3-folds (possibly with extra branes).
In both cases, the computation is usually done by decomposing a 3-manifold (resp. a Calabi-Yau 3-fold)
into elementary pieces, for which the invariants are readily available.
As basic building blocks, one can take {\it e.g.} tetrahedra and patches of $\C^3$, respectively.
Indeed, just like 3-manifolds can be built out of tetrahedra,
Calabi-Yau 3-folds can be constructed by gluing local patches of the $\C^3$ geometry.
For this reason, a tetrahedron might be called the ``simplest 3-manifold,''
whereas $\C^3$ might be called ``the simplest Calabi-Yau.''

Furthermore, in both cases the invariants associated to these basic building blocks
involve dilogarithm functions (classical and quantum). In quantum topology,
the quantum dilogarithm is the $SL(2)$ invariant associated to an ideal tetrahedron, from
which one can construct partition function of $SL(2)$ Chern-Simons theory on a generic 3-manifold \cite{DGLZ,Tudor}.
Similarly, the partition function of a local toric Calabi-Yau 3-fold (with branes) can be computed by
gluing several copies of the {\it topological vertex} associated to each $\C^3$ patch \cite{AKMV}
and, in the most basic case, the answer reduces to the quantum dilogarithm function.

\begin{figure}[ht]
\centering
\includegraphics[width=0.3\textwidth]{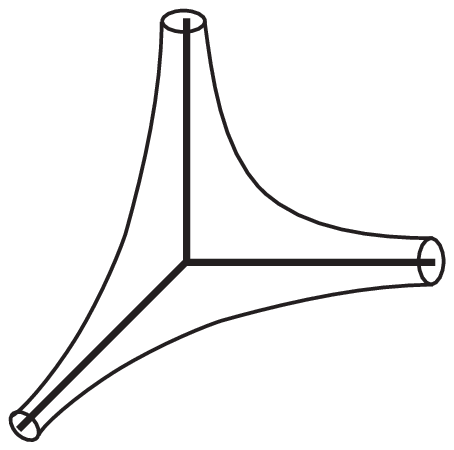}
\caption{Mirror curve for $\C^3$ geometry.}
\label{fig:mirrorC3}
\end{figure}

As in many other examples discussed in this paper, the exact solution to both
of these problems is determined by a quantization of a classical algebraic curve.
The complex curve associated to an ideal tetrahedron is simply the zero locus of
the $A$-polynomial $A(x,y) = 1 + x + y$, {\it cf.} section \ref{sec:knots}.
In topological string theory, this is the mirror curve for the $\C^3$ geometry.
More precisely, there is a whole family of such curves labeled by the so-called framing parameter $f$,
such that\footnote{One can invert the curve equation \cite{BKMP,vb-ps} to find the  expansion
$y(x) = -1 + \sum_{k=1}^{\infty} (-1)^{k(f+1)} \frac{(-k f + k - 2)!}{(-k f - 1)! k!} x^k$
(where the factorial function with negative argument is understood as the appropriate $\Gamma$-function).}
\be
A(x,y) = 1 + y + x y^f \,,
\label{Hxy}
\ee
where $x\,, y \in\C^*$ and, as usual, $x=e^u$ and $y=e^v$.
The curve \eqref{Hxy} can be visualized by thickening the edges of the toric diagram of $\C^3$, as shown in figure \ref{fig:mirrorC3}.
Various choices of framing are related by symplectic transformations $(x,y)\mapsto(xy^f,y)$,
under which closed string amplitudes $F_g$ are invariant, while $W^g_n$ and $S_k$ transform as discussed in section \ref{sec:choices}.

For integer values of $f$, the curve (\ref{Hxy}) is an $f$-sheeted cover of the $x$-plane.
There are various possible choice of parametrization of this curve,
which can be related by Galois transformations.
In following subsections, we find the corresponding quantum curves from several perspectives.
First, in subsection \ref{ssec-C3f}, we choose one very natural parametrization
and determine the corresponding quantum curve for arbitrary $f$.
Then, in subsection \ref{ssec-C3f2} we set $f=2$ and demonstrate how the form of the quantum curve changes under a change of parametrization.
Finally, in subsection \ref{ssec-C3f01}, we discuss some special choices of framing, $f=0$ and $f=1$, for which the topological recursion
cannot be applied directly, but the answer can nevertheless be obtained by treating $f$ as a continuous parameter
and considering limits of our results for generic $f$.

\bigskip


\subsection{General framing}
\label{ssec-C3f}

We wish to find a quantum version of the curve defined by the zero locus of \eqref{Hxy},
\be
\cC: \quad 1 + y + x y^f \; = 0 \,.
\label{Hxycurve}
\ee
As we explained earlier, the answer depends on the choice of parametrization.
Here we make the most convenient choice
\be
\left\{\begin{array}{l} u(p) = \log \frac{-1-p}{p^f}  \\
v(p) = \log p \end{array}\right.
\label{C3f-param}
\ee
such that $x(p)$ and $y(p)$ are rational functions.
As one can easily verify, these rational functions have trivial tame symbols \eqref{tamesymb} at all points $p \in \cC$,
which means \cite{MilnorK} that our K-theory criterion \eqref{quantcritK2} is automatically satisfied
and the curve \eqref{Hxycurve} should be quantizable for all values of $f$.

In fact, we can immediately make a prediction for what the form of the quantum curve should be,
by writing the classical curve \eqref{Hxy} in the form $A(x,y)=x+P(y)$, with $P(y)=(1+y)y^{-f}$.
This is the same form as we considered in (\ref{Alinear}),
and the parametrization \eqref{C3f-param} is consistent with the one in \eqref{xPy-param}.
Therefore, \eqref{xPy-Ahat} implies that the quantization of \eqref{Hxy}~is
\be
\widehat{A} = 1+q^{-1/2} \hat{y} + q^{(f+1)/2} \hat{x} \hat{y}^f \,.    \label{Ahat-C3f}
\ee
This result, however, is based only on the first quantum correction (\ref{A1x}) and the assumption
that all higher-other corrections can be summed up into factors of $q$.
Now we wish to show that this is indeed the case by a direct analysis of the higher order corrections.



Our first task is to determine the subleading terms $S_n$ in the wave-function (\ref{Zpert}) associated to the curve (\ref{Hxycurve}).
In order to use the topological recursion, we first need to find the branch points.
Solving the equation (\ref{branchpoint}) we find a single branch point at
\be
x_* = -\frac{f^f}{(1-f)^{1-f}} \,,\qquad y_* = p_* = \frac{f}{1-f} \,.   \label{ystar}
\ee
Note, this result is our first hint that special values of framing $f=0,1$ require extra care:
one can not simply set $f=0$ or $f=1$ from the start since for those values \eqref{ystar} gives $y_*\notin \C^*$.
In these exceptional cases, our strategy will be to carry out all computations with $f$ generic,
and then set $f=0$ or $f=1$ only in the final expressions.

The next ingredient we need is the location of the conjugate point $\overline{p}$ introduced in (\ref{conjugate}).
For the above curve, the value of $\overline{p}$ cannot be found in closed form. However, if we write
\be
p = p_* + r \,,    \label{ystarp}
\ee
we can find the conjugate point as a power series expansion in a local coordinate $r$
$$
\overline{p} = p_* -r + \frac{2(1-f^2)r^2}{3f} -  \frac{4(1 - f^2)^2 r^3}{9f^2}
+ \frac{2(1-f)^3 (22 + 57 f + 57 f^2 + 22 f^3) r^4}{135 f^3} + \mathcal{O}(r^5) \,.
$$

The remaining ingredients of the recursion are the following.
Because the curve~\eqref{Hxycurve} has genus zero, the Bergman kernel is given by a simple formula~\eqref{Bergman}.
We also find $\omega$ and $dE_q(p)$ and hence determine the recursion kernel~\eqref{kernel}.
Using local coordinates $q$ and $r$ centered at the branch point (\ref{ystarp}),
the recursion kernel has a $q$-expansion that starts with
\bea
K(q,r) & = & \frac{f^2}{2 (1 - f)^4 r^2\,q} + \frac{f (1 + f)}{2 (1 - f)^3 r^2} +  \nonumber \\ 
& & + \frac{f \big(2 f^2 r (-1 + 2 r) + 2 r (1 + 2 r) + f (3 - 8 r^2)\big) \,q}{ 6 (1 - f)^4 r^4} + \mathcal{O}(q^2) \,.  \nonumber
\eea
Even though we do not make much use of the anti-derivative, 
we mention that it can be found in the exact form,
$$
S_0 (r) = -\frac{f}{2} \log\Big(r + \frac{f}{1-f}\Big)^2 +  \log\Big(r + \frac{f}{1-f}\Big) \log\Big(\frac{1+(1-f)r}{1-f}\Big) +
 \Li_2\Big( \frac{f + (1 - f) r}{-1 + f}\Big),
$$
expressed in a local coordinate $r$, {\it cf.} \eqref{ystarp}.

Using all these ingredients, the topological recursion leads to the following results for the amplitudes~\eqref{Agn-p}:
\bea
S_2 & = & - \frac{f^2 \big(-3 + (-1 + f) f\big) + (-1 + f) f \big(3 + f (-3 + 2 f)\big) y + (-1 +
    f)^4 y^2}{24 (-1 + f) \big(f + (-1 + f) y\big)^3}, \nonumber \\
S_3 & = & \frac{f y (1 + y) \big(-2 +
   8 f^2 - (-1 + f) (1 + y) (2 - 2 y +
      f (2 + 7 y + f (2 - 7 y + 2 f (1 + y))))\big)}{48 \big(f + (-1 + f) y\big)^6},  \nonumber
\eea
and so on. We again stress that we obtain these amplitudes in a closed form, defined globally on the curve.
Now, in turn, we can apply the hierarchy of equations (\ref{hierarchy-SA}) to determine corrections $\widehat{A}_k$
to the curve~\eqref{Hxy}. To this end, we also need the following derivatives
\bea
\frac{dy}{dx}  & = & - \frac{y^{1+f}}{y + f x y^f} \,, \nonumber \\
\frac{d^2y}{dx^2} & = & \frac{f y^(1 + 2 f) (2 y + (1 + f) x y^f)}{(y + f x y^f)^3} \,, \nonumber
\eea
{\it etc.} Substituting the leading (\ref{def-Phi}) and the subleading (\ref{A1x}) terms
\bea
x\partial_x S_0 & = &  \log y \,,   \nonumber \\
S_1 & = & -\frac{1}{2}\log \frac{y-f-fy}{y(y+1)} \,,   \nonumber
\eea
into the hierarchy (\ref{hierarchy-SA}) we find the first few quantum corrections
\bea
\widehat{A}_1 & = & -\frac{1}{2}  (1 + f + 2 \hat{y} + f \hat{y}) \,, \nonumber \\
\widehat{A}_2 & = & \frac{1}{8} \big((1 + f)^2 + (2 + f)^2 \hat{y} \big) \,, \nonumber \\
\widehat{A}_3 & = & -\frac{1}{48} \big((1 + f)^3 + (2 + f)^3 \hat{y} \big) \,. \nonumber
\eea
These corrections clearly arise from the $\hbar$-expansion of $e^{-(f+1)\hbar/2 } + e^{-(1+f/2)\hbar} y + x y^f.$
Equivalently, choosing a slightly different overall normalization constant, the quantum curve reads
\be
\widehat{A} = 1+q^{-1/2} \hat{y} + q^{(f+1)/2} \hat{x} \hat{y}^f \,,
\nonumber
\ee
in perfect agreement with the original prediction~\eqref{Ahat-C3f}.

Let us mention that one can also compute from the topological recursion the coefficients $F_g$ defined in (\ref{Fg}).
As shown in \cite{vb-ps}, for the mirror $\C^3$ curve this leads to the $\hbar$-expansion
of the square root of the MacMahon function, in agreement with the closed topological string free energy.
For more complicated toric manifolds (like generalized conifolds analyzed in section \ref{sec-conifold})
the corresponding constant contributions to the (closed) partition functions
turn out to be given by multiplicities of the MacMahon function.
They are also reproduced by the topological recursion computation,
which in this case can be interpreted in terms of a pant decomposition of the mirror curve,
and mirrors A-model localization computation \cite{vb-ps}.

We can also demonstrate that the form of the above quantum curve is
consistent with, and annihilates the B-brane partition function in the
topological string theory, if conventions are adjusted appropriately.
The B-brane partition function, in arbitrary framing $f$, in the
topological vertex formalism, can be represented as\footnote{We
shifted the argument $x$ by $q^f$ to match our conventions with the
topological vertex ones. Also note, that for framing $f$, one has
$$
\langle \Tr U^m \rangle \; = \; \frac{[m + f m -1]!}{m [f m]! [m]!} \,,
$$
where $[x] = q^{x/2} - q^{-x/2}$ is the $q$-number.
Notice that for $f=0$ it reduces to $\frac{1}{m [m]}$,
which is the answer for zero framing leading to the dilogarithm.
We do not know a product formula for
$$
\sum_{m=1}^{\infty} \frac{[m + f m -1]!}{m [f m]! [m]!} x^m \, .
$$
} 
\begin{align}
\psi_f (x\, q^f) &:= \sum_{\mu} (-1)^{f |\m |} e^{\frac{f}{2} \hbar
\kappa (\mu^t)} s_{\mu^t}(x\,
q^f)\,C_{\phi\phi\mu}(e^{\hbar},e^{\hbar}) \notag \\
 &= \sum_{\m=0}^\infty \frac{(-1)^{(f+1) \m} e^{\frac{f}{2} \hbar \m (\m - 1)}
 e^{\hbar \m (f+1/2)}x^\m}{(1-e^{\hbar})\cdots(1-e^{\m \hbar})}
 = \sum_{\m=0}^\infty \frac{(-1)^{(f+1) \m} q^{\frac{\m}{2} +
\frac{f}{2} \m (\m + 1)}
 x^\m}{(1-q)\cdots(1-q^{\m})} \,,
\end{align}
where $|\mu |$ is the total number of boxes in the partition $\mu$. As
the Schur function $s_{\mu^t}$ with a single argument forces
partitions involved to be effectively one-dimensional, in the second
line we changed the domain of summation to integers. Also note that a
general expression
\be
\kappa (\mu) = |\mu | + \sum_i (\mu_i^2 - 2 i \mu_i)
\ee
in our case gives
\be
\kappa (\mu) = \m + \sum_{i=1}^{\m} (1 - 2i) = - \m (\m -1)
\ee
and $\kappa (\mu^t) 
= \m (\m - 1)$.
The function $\psi_f $ can be interpreted as  a framed invariant of
the unknot on the three-sphere. Let us now write $\psi_f (x\, q^f) =
\sum_{\m=0}^\infty a_{\m}$, with
\be
a_{\m} \; = \; \frac{(-1)^{(f+1) \m} q^{\frac{\m}{2} + \frac{f}{2} \m (\m + 1)}
x^\m}{(1-q)\cdots(1-q^{\m})} \,.
\ee
Then,
\be
\frac{a_{\m+1}}{a_{\m}} \; = \;
- x \frac{(-1)^{f} q^{\frac{1}{2} + f (\m+1)}}{(1-q^{\m+1})} \,,
\ee
so that
\be
(1-q^{\m+1}) a_{\m + 1} = -x (-1)^{f} q^{\frac{1}{2} + f (\m+1)} a_{\m}
\ee
Summing over $\m$, we get
$$
\left( 1 - \hat y + q^{f+1/2} \hat{x}  (-\hat y)^{f} \right) \psi_f
(x\, q^f) = 0.
$$
As we stressed before, there is a freedom of shifting the subleading
$S_1$ term in the partition function by a linear term in $u$. To match
to our conventions we define $Z_f(x)=x^{1/2}\psi_f(x\, q^f)$, and
commuting the additional $x^{1/2}$ in the above equation we find that
\be
\left( 1 + q^{-1/2}(-\hat y) + q^{(f+1)/2} \hat{x}  (-\hat y)^{f}
\right) Z_f (x) = 0. \nonumber 
\ee
Therefore, up to a sign of $\hat{y}$ which also is a matter of
convention, we reproduce the quantum curve which we found in
(\ref{Ahat-C3f}) in our formalism.


\subsection{Framing $f=2$}    \label{ssec-C3f2}

So far we discussed mirror curve for $\C^3$ geometry in an arbitrary framing $f$, but with a special choice of parametrization.
Now we do roughly the opposite, and discuss how the form of the quantum curve depends on the choice
of parametrization, but with a particular choice of framing $f=2$,
\be
A(x,y) = 1 + y + xy^2 \,.    \label{A-C3f2}
\ee
This curve has two branches $y^{(\alpha)}$ labeled by $\alpha=\pm$, such that
\be
y^{(\pm)} = \frac{-1\pm\sqrt{1-4x}}{2x} \,.  \nonumber 
\ee
We note that these two branches are mapped to each other by the Galois transformation
\be
x \mapsto x \,, \qquad y \mapsto \frac{1}{xy}
\label{y-xy}
\ee
that preserves the form of the curve~\eqref{A-C3f2}. From the equation of the curve we also have
\bea
\frac{dy}{dx} & = &  -\frac{A_x}{A_y} = -\frac{y^2}{1+2xy} \,,\nonumber\\
\frac{d^2y}{dx^2} & = & 2\frac{A_x A_{xy}}{A_y^2} - \frac{A_{xx}}{A_y} - \frac{A_x^2 A_{yy}}{A_y^3} = \frac{2y^3 (2+3xy)}{(1 + 2xy)^3} \,, \label{AC3f2prims} \\
\frac{d^3 y}{dx^3} & = & - \frac{6y^4 (5 + 14xy + 10x^2 y^2) }{(1 + 2xy)^5} \,. \nonumber
\eea


\subsubsection{Topological recursion}

Let us apply the topological recursion to the curve (\ref{A-C3f2}). We will consider two different parametrizations related by the symplectic transformation (\ref{y-xy}). The first parametrization which we consider is the natural one
\be
\left\{\begin{array}{l} u(p) = \log x(p) = \log\frac{-1-p}{p^2}\\
v(p) = \log y(p) = \log p \end{array}\right.   \label{C3f2-param}
\ee
It leads to a single branch point $dx(p_*)=0$ with $p_*=-2$. The conjugate of a point $p$ is
$$
\overline{p} = -\frac{p}{1+p} \,.
$$
The recursion kernel \eqref{kernel} and the anti-derivative \eqref{def-Phi} can be found in the closed form
(here we use a local parameters $q,r$, defined such that $p=p_*+q$):
\bea
K(q,r) & = & \frac{(2 - q)^2 (q-1)}{2 \big(q^2 (-1 + r) + r^2 - q r^2\big) \log(1-q)} \,, \nonumber \\
S_0 (q) & = & \log(q-2)\, \log\big(\frac{q-1}{q-2}\big) + \textrm{Li}_2(2-q) \,. \nonumber
\eea
Computing the annulus amplitude and solving the topological hierarchy we find
\bea
S_1 & = & -\frac{1}{2} \log \frac{2 + y}{x y^3}\,, \nonumber \\
S_2 & = & \frac{4 - 10 y - y^2}{24 (2 + y)^3}\,, \nonumber \\
S_3 & = & -\frac{5 y^2 (1 + y)}{4 (2 + y)^6}\,,   \label{Sk-C3f2param1} \\
S_4 & = & \frac{y (1 + y) (4096 +  y (8448 + y (-22592 + y (-25344 + y(5122 + y(162 + 7 y))))))}{5760 (2 + y)^9}.  \nonumber
\eea
Computing derivatives and using the results (\ref{AC3f2prims}), we get
\bea
S'_1 & = & \frac{1}{2} -\frac{xy(3+y)}{(2+y)(1 + 2xy)}\,, \nonumber \\
S'_2 & = & -\frac{x y^2 (-32 + 16 y + y^2)}{24 (2 + y)^4 (1 + 2 x y)}\,, \nonumber \\
S'_3 & = & -\frac{(5 x y ^3 (-4 - 2 y + 3 y^2)}{4 (2 + y)^7 (1 + 2 x y)}\,,   \label{Sk-C3f2param1prim} \\
S'_4 & = & -\frac{x y^2 (8192 + 17408 y - 172672 y^2 - 298624 y^3 + 37460 y^4 +
    144296 y^5 - 13486 y^6 - 226 y^7 - 7 y^8)}{ 5760 (2 + y)^{10} (1 + 2 x y)}.  \nonumber
\eea

Now, let us consider another parametrization, which is related to~\eqref{C3f2-param}
by the transformation $y \to (xy)^{-1}$ given in (\ref{y-xy}), so that
\be
\left\{\begin{array}{l} u(p) = \log x(p) = \log\frac{-1-p}{p^2}\\
v(p) = \log y(p) = \log \frac{-p}{p+1} \end{array}\right.
\label{C3f2-param2}
\ee
In this parametrization the equation (\ref{A-C3f2}) is also satisfied.
Since we did not redefine $x$, the expressions for the branch point $p_*=-2$
and for the conjugate $\overline{p} = -p/(1+p)$ of a point $p$ are still the same as in the previous parametrization.
The recursion kernel and the anti-derivative in the present case read (again, using local coordinates $q$ and $r$ vanishing at the branch point):
\bea
K(q,r) & = & \frac{(2 - q)^2 (1 - q)}{2 \big(q^2 (-1 + r) + r^2 - q r^2\big) \log(1-q)}\,, \nonumber \\
S_0 (q) & = & -\big(\log(q-2)\big)^2  + \frac{1}{2} \log(q-1)\, \log\big(\frac{(q-2)^2}{q-1}\big) - \textrm{Li}_2(2-q)\,. \nonumber
\eea
Using the new parametrization we compute the annulus amplitude and solve topological hierarchy to find
\bea
S_1 & = & -\frac{1}{2} \log \frac{-(1 + y)^2 (2 + y)}{x y^3}\,, \nonumber \\
S_2 & = & -\frac{(1 + y) (4 + 18y + 13 y^2)}{24 (2 + y)^3}\,, \nonumber \\
S_3 & = & -\frac{5 y^2 (1 + y)^3}{4 (2 + y)^6}\,,   \label{Sk-C3f2param2} \\
S_4 & = &  \frac{y (1 + y) (4096 + y (16128 +
    y (-3392 + y (-67584 + y (-77438 + 13 y (-1686 + 259 y))))))}{5760 (2 + y)^9}.  \nonumber
\eea
Finally, computing derivatives we get
\bea
S'_1 & = & - \frac{xy(3+2y)}{(2+3y+y^2)(1 + 2xy)}, \nonumber \\
S'_2 & = &  \frac{x y^2 (32 + 80 y + 47y^2)}{24 (2 + y)^4 (1 + 2 x y)}, \nonumber \\
S'_3 & = & \frac{5 x y^3 (1+y)^2 (4 + 6 y - y^2)}{4 (2 + y)^7 (1 + 2 x y)},   \label{Sk-C3f2param2prim} \\
S'_4 & = & \frac{x y^2 \, f_4(x,y)}{5760 (2 + y)^{10} (1+2xy)},  \nonumber \\
& & \textrm{where} \quad f_4(x,y) = -8192 - 48128 y + 65152 y^2 + 644224 y^3 + 1095340 y^4 + \nonumber \\
& & \qquad \qquad + 612184 y^5 -  38354 y^6 - 90974 y^7 + 3367 y^8. \nonumber
\eea
Not surprisingly, the perturbative coefficients \eqref{Sk-C3f2param1} and \eqref{Sk-C3f2param2}
are different in two different parametrizations that we have considered.
However, one can immediately check that they are, in fact, related by the transformation~\eqref{y-xy}.
Therefore, as expected, the entire partition function $Z$ also enjoys the action of~\eqref{y-xy}.


\subsubsection{Quantum curves}

Once we found the coefficients $S'_k$ of the perturbative expansion, we can plug our results
into the hierarchy (\ref{hierarchy-SA}) to produce the quantum corrections $\widehat{A}_k$ and,
hence, the entire quantum curve $\widehat{A}$. As usual, we start with the leading term
\be
S'_0 = \log y \,,   \label{S0prim-C3f2}
\ee
which is the same in both parametrizations, and then use higher order amplitudes computed above.
We start with the first parametrization~\eqref{C3f2-param}, in which the derivatives of $S_k$ summarized in~\eqref{Sk-C3f2param1prim}.
From the hierarchy of equations (\ref{hierarchy-SA}) we get
\bea
\widehat{A}_1 & = & - \Big(\frac{S''_0}{2}\partial_v^2 + S'_1 \partial_v \Big) A_0  = -\frac{3}{2} - 2\hat{y} \,,  \nonumber \\
\widehat{A}_2 & = & \frac{9}{8} + 2 \hat{y}\,, \nonumber \\
\widehat{A}_3 & = & -\frac{9}{16} -\frac{4}{3} \hat{y}\,.   \nonumber
\eea
These coefficients arise from the $\hbar$-expansion of $e^{-3\hbar/2} + e^{-2\hbar}\hat{y}+\hat{x}\hat{y}^2$ and,
therefore, up to an overall normalization, the quantum curve \eqref{Ahatpert} in this case reads
\be
\widehat{A}(\hat{x},\hat{y}) = 1 + q^{-1/2}\hat{y} + q^{3/2} \hat{x} \hat{y}^2 \,,
\label{Ahat-C3f2param1}
\ee
in agreement with (\ref{Ahat-C3f}) for $f=2$.

We can also consider the second parametrization \eqref{C3f2-param2}.
The leading term $S'_0$ is the same as~\eqref{S0prim-C3f2},
and the higher order perturbative corrections are given by~(\ref{Sk-C3f2param2prim}).
This time, the hierarchy \eqref{hierarchy-SA} leads to
\bea
\widehat{A}_1 & = & -\frac{3}{2} - \hat{y} \,,    \nonumber \\
\widehat{A}_2 & = & \frac{9}{8} + \frac{\hat{y}}{2} \,, \nonumber \\
\widehat{A}_3 & = & -\frac{9}{16} - \frac{\hat{y}}{6} \,.   \nonumber
\eea
These terms (up to an overall normalization) arise from the expansion of the quantum curve
\be
\widehat{A}(\hat{x},\hat{y}) = 1 + q^{1/2} \hat{y} + q^{3/2} \hat{x} \hat{y}^2,
\label{Ahat-C3f2param2}
\ee
which is different from (\ref{Ahat-C3f2param1}).

Finally, the present example gives us a good opportunity
to illustrate how the factorization (\ref{Ahat-factor})
works for curves in $\C^*\times\C^*$.
Indeed, it is easy to see that to the leading order in $\hbar$ the quantum curve factorizes as
\be
\widehat{A} = 1 + \hat{w} - \frac{p+1}{p^2}\hat{w}^2 +
\mathcal{O}(\hbar) = (p-\hat{w})(p + (p+1)\hat{w}) +
\mathcal{O}(\hbar) \,,
\label{factorizedii}
\ee
where we used (\ref{C3f2-param}) and also introduced $\hat{w}=e^{-\frac{p(p+1)}{p+2}\hbar \partial_p}$.
In this factorized expression, the first factor $(p-\hat{w})$ annihilates the wave function
$$
Z = e^{-\frac{1}{\hbar}\int dp \frac{p+2}{p(p+1)}\log p} \Big(1 +
\mathcal{O}(\hbar)\Big) = e^{\frac{1}{\hbar}\big(
\textrm{Li}_2(-p)+\log p \cdot \log(1+p^{-1}) \big)} \Big(1 +
\mathcal{O}(\hbar)\Big) \,.
$$
The exponent here indeed reproduces the leading order term in the partition function,
$S_0=\int v(p) d u(p)$, in the parametrization (\ref{C3f2-param}). On the other hand,
{}from the second factor $p+(p+1)\hat{w}$ in \eqref{factorizedii} one finds $S_0$ in the second parametrization~\eqref{C3f2-param2}.


\subsection{Framing $f=0,1$}
\label{ssec-C3f01}

In the preceding subsections, we found the quantum curves for a tetrahedron (or $\C^3$) model with a generic framing,
and also analyzed in excruciating detail the case $f=2$.
The situation becomes more delicate for special values of framing $f=0,1$ because in these cases
the branch point (\ref{ystar}) escapes ``to infinity'' and the topological recursion can no longer be directly applied.
However, as also stressed in \cite{vb-ps}, one can still obtain meaningful results by treating $f$ as a continuous parameter, and taking the limit $f\to 0,1$ in the end of the computation.

Let us analyze the case $f=0$ from this viewpoint first. From the general result (\ref{Ahat-C3f})
we conclude that for $f=0$ the quantum curve should take the form
\be
\widehat{A}_{f=0} \; = \; 1 + q^{-1/2}\hat{y} + q^{1/2}\hat{x} \,.
\label{Ahat-C3f0}
\ee
The partition function $Z$ associated to this operator is given by a version
of the quantum dilogarithm (\ref{dilog}) and can be written as
\be
Z_{f=0} \; = \; c\cdot x^{1/2} \psi(-x) \,,
\label{ZC3f0}
\ee
where $c$ is some multiplicative factor which is not fixed by the $q$-difference equation \eqref{AhatZ0}.
This form of the partition function follows from the application of the differential hierarchy (\ref{hierarchy-SA})
to the quantum curve (\ref{Ahat-C3f0}), or can be seen directly as follows.
Assuming that the constant normalization factor $c$ contains $\prod_k (-1)=(-1)^{\zeta(0)}$
and changing the signs in each factor of the product (\ref{dilog}) we see that
\be
\hat{y} Z_{f=0}  = q^{1/2} x^{1/2}\prod_{k=1}^{\infty} \frac{1}{-1-x q^{k+1/2}} = q^{1/2} (-1-x q^{1/2}) Z_{f=0} \,,  \label{ZC3f0-derive}
\ee
which is equivalent to the statement $\widehat{A}_{f=0}  Z_{f=0}  = 0$.

Now, let us compare the perturbative $\hbar$-expansion of the partition function (\ref{ZC3f0})
with what one might find from the topological recursion. The leading term is
$$
S_0 = \int \frac{\log(-1-x)}{x} dx = i\pi\log x - \textrm{Li}_2(-x) \,,
$$
where the dilogarithm properly reproduces the leading term in (\ref{dilog}).
The next, subleading contribution given by the annulus amplitudes is
$$
S_1 = \frac{i\pi}{2} + \frac{1}{2}\log x \,,
$$
and, again, it reproduces the corresponding factor $x^{1/2}$ in (\ref{ZC3f0}).
The higher order terms $S_k$ arise from the topological recursion as follows.
First, notice that all $W^g_n$ with $n\neq 1$ vanish for $f=0$.
This immediately implies that all $S_{2k+1}=0$ because only $W^g_n$ with even values of $n$ contribute to $S_{2k+1}$.
On the other hand, the correlators with $n=1$, which remain non-zero in the $f\to 0$ limit, read
\bea
W^1_1(p) & = & \frac{1}{24p^2} \,, \nonumber \\
W^2_1(p) & = & -\frac{7(6+6p+p^2)}{5760 p^4} \,, \nonumber \\
W^3_1(p) & = & \frac{31(120+240p+150p^2+30p^3+p^4)}{967680 p^6} \,. \nonumber
\eea
Integrating these correlators (and including an appropriate integration constant in $S_2$) we find
the following functions of $x$,
\bea
S_2 & = & \frac{1}{24}\textrm{Li}_0(-x) \,, \nonumber \\
S_4 & = & -\frac{7}{5760} \textrm{Li}_{-2}(-x) \,, \nonumber \\
S_4 & = & -\frac{31}{967680} \textrm{Li}_{-4}(-x) \,, \nonumber
\eea
which, as expected, agree with the expansion (\ref{qdilog-hbar}).
In topological string theory, this partition function represents a $B$-brane amplitude in the $\C^3$ geometry.

In the second special limit, $f\to 1$, the situation is a little more subtle due to the divergence of the correlators $W^g_{2k}$.
This, however, does not affect the leading terms $S_0$ and $S_1$ which still can be computed by direct methods.
The higher-order terms, on the other hand, can be obtained from the hierarchy of equations (\ref{hierarchy-SA})
applied to the quantum curve (\ref{Ahat-C3f}) with $f=1$:
\be
\widehat{A}_{f=1}  = 1 + q^{-1/2}\hat{y} + q \hat{x} \hat{y} \,.
\label{Ahat-C3f1}
\ee
{}From the topological string point of view, this choice of framing corresponds to an anti-$B$-brane,
whose partition function should be roughly the inverse of that for a $B$-brane.
Curiously, however, the hierarchy (\ref{hierarchy-SA}) applied to the above quantum curve
reveals that the $\hbar$-expansion of the free energy contains not only polylogarithms of even order,
but also polylogarithms of odd order. This expansion starts with
$$
S_0=\textrm{Li}_{2}(-x) \,,\qquad   S_1 
= \log x^{1/2} + \textrm{Li}_{1}(-x) \,,\qquad
S_2=\frac{11}{24}\textrm{Li}_{0}(-x) \,,\qquad
S_3=\frac{1}{8} \textrm{Li}_{-1}(-x) \,,
$$
and can be summed up to a generating function
$$
Z_{f=1} = \frac{c\cdot x^{1/2}}{\psi(-x)} e^{\sum_{k=0}^{\infty} \frac{\hbar^k}{2^k k!} \textrm{Li}_{1-k} (-x) } =
\frac{c \cdot x^{1/2}}{\psi(-x)} e^{-\log(1+ x e^{\hbar/2})} = c \cdot x^{1/2} \prod_{k=1}^{\infty} \big(1 + x e^{\hbar(k+1/2)} \big).
$$
As a check of this result we make an observation analogous to (\ref{ZC3f0-derive}):
$$
\hat{y} Z_{f=1}  = q^{1/2} x^{1/2}\prod_{k=1}^{\infty} \big( -1-x q^{k+3/2} \big) = q^{1/2} \frac{ Z_{f=1} }{-1-x q^{3/2}} \,,
$$
where we also identified the multiplicative factor $c$ with $\prod_k (-1)=(-1)^{\zeta(0)}$.
After multiplying both sides of this expression by the denominator $1+x q^{3/2}$ we recover the quantum curve equation (\ref{Ahat-C3f1}).




\section{Conifold and generalizations}
\label{sec-conifold}

\begin{figure}[ht]
\centering
\includegraphics[width=0.7\textwidth]{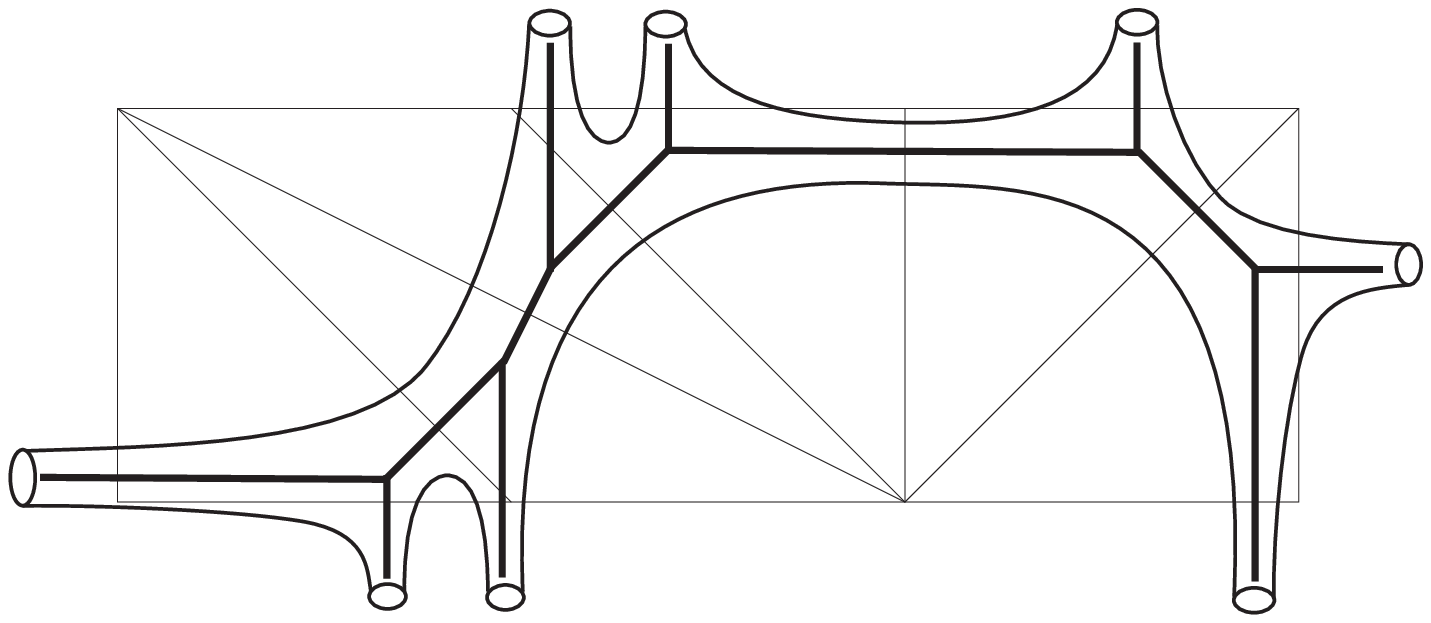}
\caption{An example of mirror curve for a generalized conifold.}
\label{fig:mirrorCurves}
\end{figure}

There is a large  class of toric Calabi-Yau manifolds, known as the generalized conifolds, whose mirror curves have genus zero.
They provide especially simple and attractive examples, for which the corresponding quantum curves can be easily determined using our technique.
Toric diagrams for this class of manifolds arise from a triangulation of a ``strip,'' as shown in figure~\ref{fig:mirrorCurves}.
The corresponding mirror curves are always linear in one of the variables. Therefore, up to a coordinate change,
they can be put in the form
\be
A(x,y) \; = \; B(x) + y C(x) \,.
\label{AgenConi}
\ee

With a suitable choice of framing, $B(x)$ and $C(x)$ can be written in a simple
product form $B(x)=\prod_i (1+Q_i x)$ and $C(x)=\prod_j (1+\widetilde{Q}_j x)$,
where $Q_i$ and $\widetilde{Q}_j$ encode the K{\"a}hler parameters of the toric Calabi-Yau 3-fold.
For this choice of framing
the partition function of generalized conifolds is always a product of quantum dilogarithms,
which can be easily recognized from the leading behavior
$$
S_0 = \int \log y \frac{dx}{x} = \Big( \sum_j \textrm{Li}_2(-\widetilde{Q}_j x) \Big) - \Big( \sum_i \textrm{Li}_2(-Q_i x) \Big) \,.
$$
The higher-order $\hbar$-corrections complete the dilogarithms here to quantum dilogarithms in the full partition function,
generalizing the expansion (\ref{qdilog-hbar}) in an obvious way.
With this particularly nice choice of framing, it is also easy to extend the computation (\ref{ZC3f0-derive})
to find corresponding quantum curves.

\bigskip

For general framing, however, a derivation of the quantum curve along these lines is by far non-obvious.
It is this point where our results turn out to be very powerful and allow to determine quantum curves
in any framing in a straightforward and systematic manner.
Writing the equation (\ref{AgenConi}) with $x$ and $y$ interchanged, as
\be
A(x,y) \; = \; B(y)+x C(y) \,,
\label{AgenConiyx}
\ee
essentially represents the same toric geometry and the same algebraic curve.
Equivalently, the curve $A(x,y)=0$ can be described as the zero locus of (\ref{Alinear}) with $P(y) = B(y)/C(y)$,
and from (\ref{xPy-Ahat}) we immediately obtain
\be
\widehat{A} = B(q^{-1/2} \hat{y}) + q^{1/2} \hat{x} \, C(q^{1/2} \hat{y}) \,.
\label{Ahat-genConi}
\ee
Because the latter choice of the generalized conifold equation (linear in $x$) differs from (\ref{AgenConi})
by the exchange of $x$ and $y$, the corresponding partition functions are related by a Fourier transform.
In particular, we mentioned earlier that for a specific choice of framing\footnote{in which $B(x)$ and $C(x)$
have a product form $B(x)=\prod_i (1+Q_i x)$ and $C(x)=\prod_j (1+\widetilde{Q}_j x)$}
the partition function $Z$ is built out of quantum dilogarithms.
Since the quantum dilogarithm is self-similar under Fourier transform,
it follows that the convolution of a product of quantum dilogarithms is again a product of quantum dilogarithms.
Hence, the Fourier transform of the partition function should also be a product of quantum dilogarithms.
This can be verified directly using the form of the quantum curve (\ref{Ahat-genConi}) and the hierarchy of equations (\ref{hierarchy-SA}).

As a check of our result (\ref{Ahat-genConi}), we note that for $B(y)=1+y$ and $C(y)=y^f$ we get
$$
\widehat{A}_{\C^3} = 1 + q^{-1/2}\hat{y} + q^{(f+1)/2} \hat{x} \hat{y}^f \,,
$$
which correctly reproduces the quantum curve (\ref{Ahat-C3f}) of the $\C^3$ geometry discussed earlier in section \ref{sec-tetrahedron}.
As another example one can consider an ordinary conifold, whose mirror curve in zero framing $f=0$ reads
$$
A_{f=0}(x,y) \; = \; 1 + x + y + Q\frac{x}{y} \,,
$$
where, as usual, $Q$ is the (exponentiated) K{\"a}hler parameter.
Similarly, for general value of framing $f$,
the mirror curve of the conifold is given by the zero locus of a degree-$f$ polynomial
\be
A_f (x,y) \; = \; 1 + x y^{f} + y + Qx y^{f-1} \,,
\label{coni-f}
\ee
which is manifestly in the form \eqref{AgenConiyx} with $B(y)=1+y$ and $C(y)=y^f+Qy^{f-1}$.
Therefore, from \eqref{Ahat-genConi} we conclude that the quantization of this $A$-polynomial is
\be
\widehat{A}_f \; = \; 1 + q^{-1/2}\hat{y} + q^{(f+1)/2} \hat{x} \hat{y}^f + Q q^{f/2} \hat{x} \hat{y}^{f-1} \,.
\label{Ahat-coni-f}
\ee
Another special choice of framing $f=2$ leads to the quantum curve (\ref{xPy-Conif2}) which will be analyzed next
to high order in topological recursion. Before we proceed to this example, however, let us remind the reader
that a particular form of the quantum curve depends not only on the classical equation but also on the choice
of parametrization, as discussed in sections \ref{sec:choices} and \ref{ssec-C3f2}, and as will be also discussed below.
For example, the quantum curves \eqref{Ahat-genConi}, \eqref{Ahat-coni-f}, and \eqref{xPy-Conif2}
all come from the choice of parametrization \eqref{xPy-param}.

Quantum curves for generalized conifolds were also studied recently in \cite{ACDKV,BDH}.
In particular, in \cite{ACDKV} a different quantization of the classical curve $A(x,y)=0$
was related to the Nekrasov-Shatashvili limit \cite{NS} of the {\it refined} topological string
partition function, where $\epsilon_1 = 0$ and $\epsilon_2 = \hbar$ (see also \cite{DGH}).
In that framework, the classical curves for generalized conifolds and even more general examples
are quantized\footnote{We thank Mina Aganagic and Robbert Dijkgraaf for clarifying discussions on this.}
by simply replacing $x$ and $y$ with $\hat x$ and $\hat y$
(where all $q$-factors in $\widehat{A}$ can be absorbed in a normalization of $\hat x$, $\hat y$, or K\"ahler parameters).
In particular, the new interesting phenomena where the numerical coefficients ``split'' into several powers of $q$,
as in
$$
A = 3x^5 + \ldots \quad \leadsto \quad \widehat{A} = (q+q^3+q^5) x^5 + \ldots
$$
or where completely new terms appear upon quantization (as in $\widehat{A} = (1-q^3) x^3 + \ldots$)
never happen in the framework of \cite{ACDKV}. It is tempting to speculate that such phenomena --- that one
encounters {\it e.g.} in quantization of $A$-polynomials for some simple knots --- can be accounted for
by going from the Nekrasov-Shatashvili limit $\epsilon_1 = 0$, $\epsilon_2 = \hbar$
to the limit $\epsilon_1 = - \epsilon_2 = \hbar$.


\subsection{Conifold in $f=2$ framing}

\begin{figure}[ht]
\centering
\includegraphics[width=0.3\textwidth]{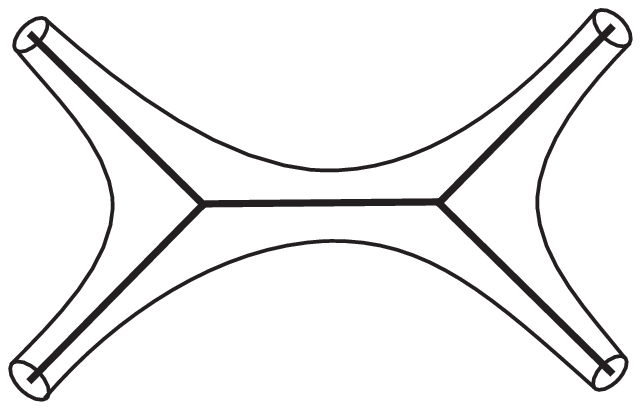}
\caption{Mirror curve for the conifold geometry.}
\label{fig:mirrorConi}
\end{figure}

In this section we analyze the ordinary conifold, whose mirror curve is shown in figure \ref{fig:mirrorConi}.
As in the case of $\C^3$ geometry, we wish to discuss a special choice of framing (namely, $f=2$)
and study how a choice of parametrization affects the form of the quantum curve.

For $f=2$, the conifold mirror curve (\ref{coni-f}) takes the form
\be
A(x,y) \equiv A_{f=2}(x,y) = 1 + y + xy^2 + Q x y \,,
\label{A-Conif2}
\ee
and in the limit $Q\to 0$ reduces to the $\C^3$ mirror curve (\ref{A-C3f2}) in the same framing.
In fact, the relation between these two models goes much further.
For example, the curve defined by the zero
locus of \eqref{A-Conif2} has two branches $y^{(\alpha)}$ labeled by $\alpha=\pm$,
\be
y^{(\pm)} = \frac{-1 - Qx \pm\sqrt{(1+Qx)^2 - 4x}}{2x} \,,
\label{Conif2-yalpha}
\ee
which, as in the $\C^3$ model, are exchanged by the Galois transformation (\ref{y-xy}):
\be
\left( x,y \right) \mapsto \left( x, \frac{1}{xy} \right) \,.   \label{coni-y-xy}
\ee
{}From the equation of the curve we also find the following formulae
\bea
\frac{dy}{dx} & = & -\frac{A_x}{A_y} = -\frac{Q y + y^2}{1+Qx+2xy} \,,     \label{AConif2prims}  \\
\frac{d^2y}{dx^2} & = & 2\frac{A_x A_{xy}}{A_y^2} - \frac{A_{xx}}{A_y} - \frac{A_x^2 A_{yy}}{A_y^3} = \frac{2y (Q+y) \big(Q + Q^2 x + (2 + 3 Q x) y + 3 x y^2\big)}{(1 + Qx + 2xy)^3} \,, \nonumber  \\
\frac{d^3 y}{dx^3} & = & - \frac{6 y (Q + y)}{ (1 + Q x + 2 x y)^5 }   \Big(Q^2 (1 + Q x)^2 +
    Q (5 + 11 Q x + 6 Q^2 x^2) y + \nonumber  \\
    & & \qquad \qquad \qquad \qquad  + (5 + 21 Q x + 16 Q^2 x^2) y^2 + 2 x (7 + 10 Q x) y^3 + 10 x^2 y^4\Big)
    .   \nonumber
\eea
which will be useful to us later.

\bigskip


\subsubsection{Topological recursion}

The curve (\ref{A-Conif2}) is quadratic and, therefore, is a double cover of the $x$-plane.
We introduce two parametrizations of this curve which, just like the two branches \eqref{Conif2-yalpha},
are permuted by the Galois transformation (\ref{coni-y-xy}).

The first parametrization is the obvious one
\be
\left\{\begin{array}{l} u(p) = \log x(p) = \log\frac{-1-p}{p(p+Q)}\\
v(p) = \log y(p) = \log p \end{array}\right.   \label{Conif2-param}
\ee
and is motivated by writing (\ref{A-Conif2}) in the form (\ref{Alinear}) with $P(y)=(1+y)/(Qy + y^2)$.
Indeed, applying our general result (\ref{xPy-Ahat}) to this particular model we immediately obtain
\be
\widehat{A} \; = \; 1 + q^{-1/2} \hat{y} + q^{3/2} \hat{x}\hat{y}^2 + q Q \hat{x}\hat{y} \,,
\label{xPy-Conif2}
\ee
which is also consistent with (\ref{Ahat-coni-f}).
As we pointed out earlier, however, this result is based only on the elementary computation
of the annulus amplitude $S_1$,
and now we wish to verify that computing $S_n$ and $\widehat{A}_n$ to higher order
does not lead to any modifications and merely confirms the result~\eqref{xPy-Conif2}.

The conifold curve (\ref{Conif2-param}) has two branch points
\be
p_* = -1 \mp \sqrt{1 - Q} \,.
\label{Conif2-brpt}
\ee
Notice, in the $Q\to 0$ limit, the branch point with the minus sign reduces to the $\C^3$ branch point $p_*=-2$,
whereas the other branch point runs away to $p_*=0\notin\C^*$.

The conjugate of a generic point $p$ is given in a global form (the same around both branch points)
$$
\overline{p} = \frac{-p-Q}{1+p} \,.
$$
The recursion kernel and the anti-derivative can be found in the closed form
\bea
K(q,z) & = & \frac{q (1 + q) (q + Q)}{2 (z-q) (q + Q + z + q z)
   \log\big(\frac{-q - Q}{q(1 + q)}\big)} \,, \nonumber \\
S_0 (q) & = &  -\frac{1}{2} \log q \Big( \log q + 2 \log\big(\frac{q+Q}{Q(1+q)}\big) \Big)
+  \textrm{Li}_2(-q)-\textrm{Li}_2(-q/Q) \,, \nonumber
\eea
{}from which we can compute the annulus amplitude and solve the topological hierarchy. We find
\bea
S_1 & = & -\frac{1}{2} \log\big(\frac{Q + y (2 + y)}{x y^2 (Q + y)^2}\big) \,, \nonumber \\
S_2 & = & \frac{y (1 - Q)  \big(11 Q^2 + 2 Q (7 - 5 y) y -
   y^2 (-4 + y (10 + y))\big)}{24 (Q + y (2 + y))^3} \,, \label{Sk-Conif2param} \\
S_3 & = & \frac{ (Q-1) y (1 + y) (Q + y) (Q - y^2) (Q^3 - 10 y^4 -
   6 Q^2 y (1 + 3 y) + Q y^2 ( y^2-26 y -6))}{8 (Q + y (2 + y))^6 }.   \nonumber
\eea

Now, let us consider another parametrization of the classical curve (\ref{A-Conif2}),
related to (\ref{Conif2-param}) by the transformation~(\ref{coni-y-xy}):
\be
\left\{\begin{array}{l} u(p) = \log x(p) = \log\frac{-1-p}{p(p+Q)}\\
v(p) = \log y(p) =  \log \frac{-p-Q}{p+1} \end{array}\right.
\label{Conif2-paramY2}
\ee
Since $x$ is not affected by the transformation~(\ref{coni-y-xy}),
we find the same two branch points (\ref{Conif2-brpt}):
$$
p_* = -1 \mp \sqrt{1 - Q} \,,
$$
whose behavior in the $Q\to 0$ limit was discussed below eq. (\ref{Conif2-brpt}).

In the new parametrization \eqref{Conif2-paramY2}, the conjugate of a point $p$
is given by the same formula as in the previous parametrization~ \eqref{Conif2-param}:
$$
\overline{p} = \frac{-p-Q}{1+p} \,.
$$
The recursion kernel and the anti-derivative can be also found in the closed form.
The kernel differs by a sign from the kernel in previous parametrization
$$
K(q,z) = \frac{q (1 + q) (q + Q)}{2 (q - z) (q + Q + z + q z)
   \log\big(\frac{-q - Q}{q(1 + q)}\big)} \,,
$$
and, as everything else, in the $Q\to 0 $ limit reduces to the recursion kernel of the $\C^3$ model.
The formula for $S_0$ can be also written explicitly, even though its form is a little involved.

Computing the annulus amplitude and solving the topological hierarchy we now find
\bea
S_1 & = & -\frac{1}{2} \log\big(\frac{(1+y)^2 (Q + y (2 + y))}{x y^2 (Q + y)^2 (Q-1)}\big) \,, \label{Sk-Conif2param2} \\
S_2 & = &  \frac{(1 + y) (Q + y) \big(Q^3 + Q^2 (1 + 2 y (7 + 5 y)) +
   y^2 (4 + y (18 + 13 y)) - Q y (6 + y (2 + y (10 + 11 y)))\big)}{24 (Q-1) (Q + y (2 + y))^3} \,,   \nonumber
\eea
which should be compared to the analogous formulae \eqref{Sk-Conif2param}
obtained in a different parametrization / polarization.


\subsubsection{Quantum curves}

Once we found the perturbative amplitudes $S_k$,
we can compute their derivatives and determine the form of the quantum curve from the hierarchy of equations (\ref{hierarchy-SA}).
With the first choice of parametrization \eqref{Conif2-param}, we get
\bea
\widehat{A}_1 & = & -\frac{\hat{y}}{2} + Q \hat{x}\hat{y} + \frac{3}{2} \hat{x} \hat{y}^2 \,, \nonumber \\
\widehat{A}_2 & = & \frac{1}{8} (\hat{y} + 4 Q \hat{x}\hat{y} + 9 \hat{x} \hat{y}^2) \,, \nonumber \\
\widehat{A}_3 & = & \frac{1}{48} (-\hat{y} + 8 Q \hat{x}\hat{y} + 27 \hat{x}\hat{y}^2) \,.   \nonumber
\eea
It is easy to see that these are precisely the coefficients which arise from the perturbative
$\hbar$-expansion of the curve~(\ref{xPy-Conif2}):
\be
\widehat{A}(\hat{x},\hat{y}) \; = \; 1 + q^{-1/2}\hat{y} + q^{3/2} \hat{x} \hat{y}^2 + q Q \hat{x} \hat{y} \,,
\ee
which, in the $Q\to 0$ limit, reduces to the quantum curve (\ref{Ahat-C3f2param1}) of the $\C^3$ model (in a similar parametrization).

In the second parametrization (\ref{Conif2-paramY2}), computing the derivatives of $S_k$ from (\ref{Sk-Conif2param2})
and substituting the result into the hierarchy of loop equations (\ref{hierarchy-SA}) gives
\bea
\widehat{A}_1 & = & -1 - \frac{\hat{y}}{2} + \frac{1}{2} \hat{x}\hat{y}^2 \,,     \nonumber \\
\widehat{A}_2 & = & \frac{1}{2} + \frac{\hat{y}}{8} + \frac{1}{8} \hat{x}\hat{y}^2 \,,    \nonumber
\eea
{\it etc.}
Up to an overall normalization,
these coefficients arise from the $\hbar$-expansion of the quantum curve
\be
\widehat{A}(\hat{x},\hat{y}) \; = \; 1 + q^{1/2}\hat{y} + q^{3/2} \hat{x} \hat{y}^2 + q Q  \hat{x} \hat{y} \,.
\ee
As expected, in the limit $Q\to 0$ this expression reduces to (\ref{Ahat-C3f2param2}).


\acknowledgments{It is pleasure to thank
Vincent Bouchard, Tudor Dimofte, Nathan Dunfield, Bertrand Eynard, Maxim Kontsevich, and Don Zagier
for helpful discussions and correspondence.
The work of S.G. is supported in part by DOE Grant DE-FG03-92-ER40701FG-02 and in part by NSF Grant PHY-0757647.
The research of P.S. is supported by the DOE grant DE-FG03-92-ER40701FG-02
and the European Commission under the Marie-Curie International Outgoing Fellowship Programme.
Opinions and conclusions expressed here are those of the authors and do not necessarily reflect the views of funding agencies.}


\appendix

\section{A hierarchy of differential equations}       \label{app:hierarchy}

In this appendix we provide more details on the hierarchy of differential equations (\ref{hierarchy-SA}) arising from the quantum curve equation $\widehat{A} Z=0$. This hierarchy allows to determine the quantum operator $\widehat{A}$, order by order in $\hbar$, from the knowledge of the partition function $Z$ it annihilates, or vice versa. We stress that the hierarchy (\ref{hierarchy-SA}) takes the same form for curves embedded in $\C\times\C$ or $\C^*\times\C^*$, even though its derivation in both cases is much different.

We recall that, in the classical limit, we consider curves embedded either in $\C\times\C$ with coordinates $(u,v)$, or in $\C^*\times\C^*$ with coordinates $(x=e^u,y=e^v)$. The classical curve is given by the polynomial equation
\be
0= A\equiv A_0.   \label{AA0}
\ee
In the quantum regime we introduce the commutation relation $[\hat{v},\hat{u}]=\hbar$ and use the representation $\hat{u}=u,\hat{v}=\hbar\partial_u$. For $\C^*$ coordinates we then have $\hat{x}=x=e^u, \hat{y}=e^{\hat{v}}=e^{\hbar\partial_u}$ and $\hat{y}\hat{x}=q\hat{x}\hat{y}$, where $q=e^{\hbar}$. In what follows we denote derivatives w.r.t $u$ by $'=\partial_u=x\partial_x$.

To represent the quantum curves corresponding to (\ref{AA0}) we use the following expansions, respectively in $\C\times\C$ and $\C^*\times\C^*$ case
$$
\widehat{A} = \sum_{j=0}^d a_j(u,\hbar) \hat{v}^j,\qquad \quad \widehat{A} = \sum_{j=0}^d a_j(x,\hbar) \hat{y}^j,
$$
where, respectively,
\be
a_j(u,\hbar) = \sum_{l=0}^{\infty} a_{j,l}(u)\hbar^l, \qquad\quad  a_j(x,\hbar) = \sum_{l=0}^{\infty} a_{j,l}(x)\hbar^l.    \nonumber 
\ee
We also reassemble contributions of fixed $\hbar$ order into, respectively,
\be
A_l= A_l(u,v) = \sum_{j=0}^d a_{j,l}(u) v^j,    \qquad\quad A_l= A_l(x,y) = \sum_{j=0}^d a_{j,l}(x) y^j.      \label{Al}
\ee
Replacing classical variables in these expansions by quantum operators $\hat{u},\hat{v}$ or $\hat{x},\hat{y}$, ordered such that $\hat{v}$ or $\hat{y}$ appear to the right of $\hat{u}$ or $\hat{x}$, defines corrections $\widehat{A}_l$ to the quantum curve (\ref{Ahatpert}).
Using the above notation, the quantum curve equation can be written, respectively in $\C\times\C$ and $\C^*\times\C^*$ case, as
\be
\widehat{A} Z(u) = \Big(\sum_{j=0}^d a_j(u,\hbar) \hat{v}^j\Big) Z(u) =  0,   \qquad \qquad
\widehat{A} Z(x) =  \Big(\sum_{j=0}^d a_j(x,\hbar) \hat{y}^j\Big) Z(x) =  0,   \label{diff-eq}
\ee
where
\be
Z = \exp\Big(\frac{1}{\hbar}\sum_{k=0}^{\infty} \hbar^k S_k  \Big).   \label{Zm}
\ee


\subsection{Hierarchy in the $\C^*$ case: $q$-difference equation}

The quantum curve equation gives rise to a hierarchy of differential equations which arise as follows. Substituting the partition function (\ref{Zm}) into (\ref{diff-eq}) and dividing by $e^{\hbar^{-1} S_0}$ results in
\be
0 = \sum_{j,l=0}^{\infty} a_{j,l} \hbar^l e^{j S'_0} \exp\Big(\sum_{n=1}^{\infty} \hbar^n \frak{d}_n(j) \Big),  \label{diff-eq-2}
\ee
where $\frak{d}_n(j)$ combine terms with a fixed power of $\hbar$ in the expansion of $\sum_k\hbar^k S_k\big(e^{u+j\hbar}\big)$
\be
\frak{d}_n(j) = \sum_{r=1}^{n+1} \frac{j^r}{r!} S^{(r)}_{n+1-r}(x).    \label{sigma-n}
\ee
For example
\bea
\frak{d}_1(j) & = & \frac{j^2}{2} S''_0 + j S'_1, \nonumber \\
\frak{d}_2(j) & = & \frac{j^3}{6}S'''_0 + \frac{j^2}{2} S''_1 + j S'_2, \nonumber \\
\frak{d}_3(j) & = & \frac{j^4}{4!}S^{(4)}_0 +\frac{j^3}{3!}S'''_1 + \frac{j^2}{2} S''_2 + j S'_3, \nonumber
\eea
and note that for each $n$ we have $\frak{d}_n(0) = 0$. Let us now expand the exponent in (\ref{diff-eq-2}) and collect terms with fixed power of $\hbar$
\be
\exp\Big(\sum_{n=1}^{\infty} \hbar^n \frak{d}_n(j) \Big) = \sum_{r=0}^{\infty} \hbar^r \D_r(j),     \label{Sigma-def2}
\ee
so that, for example,
\bea
\D_0(j) & = & 1, \nonumber \\
\D_1(j) & = & \frak{d}_1(j) = \frac{S''_0}{2}  j^2 + S'_1 j ,  \nonumber \\
\D_2(j) & = & \frak{d}_2(j) + \frac{1}{2}\frak{d}_1(j)^2 = \frac{(S''_0)^2}{8}   j^4+  \frac{1}{6}\big(S'''_0 + 3S''_0 S'_1 \big)j^3 + \frac{1}{2}\big(S''_1 + (S'_1)^2 \big)j^2 + S'_2 j,  \nonumber \\
\D_3(j) & = & \frak{d}_3(j) + \frak{d}_1(j) \frak{d}_2(j) + \frac{1}{6} \frak{d}_1(j)^3 =    \nonumber \\
& = & \frac{(S''_0)^3}{48} j^6 + \Big(\frac{S''_0 S'''_0}{12} + \frac{(S''_0)^2 S'_1}{8}  \Big)j^5 +
\frac{1}{24} \big(S''''_0 + 6 S''_0 S''_1 + 4 S'''_0 S'_1 + 6 S''_0 (S'_1)^2 \big) j^4 + \nonumber \\
& & +  \frac{1}{6} \big(3 S''_1 S'_1 + (S'_1)^3 + S'''_1 + 3 S''_0 S'_2\big) j^3 +
\big(\frac{S''_2}{2} + S'_1 S'_2) j^2 + S'_3 j , \nonumber\\
\D_4(j) & = & \frak{d}_4(j) + \frak{d}_1(j) \frak{d}_3(j) + \frac{1}{2} \frak{d}_2(j)^2 + \frac{1}{2} \frak{d}_1(j)^2 \frak{d}_2(j) + \frac{1}{4!}\frak{d}_1(j)^4 =   \nonumber \\
& = & \frac{(S''_0)^4}{384} j^8 + \frac{1}{48} \big( (S''_0)^2 S'''_0 + (S''_0)^3 S'_1 \big) j^7 + \ldots + \frac{1}{2}\big((S'_2)^2 + S''_3 + 2S'_1 S'_3  \big)  j^2 + S'_4 j.\nonumber
\eea
Finally, expanding (\ref{diff-eq-2}) in total power of $\hbar$ and collecting terms with a fixed such power $\hbar^n$, gives rise to a hierarchy of differential equations
\be
0 = \sum_j e^{jS'_0} \sum_{r=0}^n a_{j,r}\D_{n-r}(j).     \label{diff-hierarchy}
\ee

Now we use the fact that the disk amplitude in $\C^*\times\C^*$ case is $S_0 =  \int \log(y) \frac{dx}{x}$, so $S' _0 = \log(y)$. Therefore $e^{jS'_0} = y^j$ and we can write (\ref{diff-hierarchy}) in terms of corrections $A_k$ to the quantum curve (\ref{Al}). In particular the first equation in the hierarchy $0=\sum_{j=0}^d a_{j,0} y^j=A_0(x,y)$ coincides with the classical curve equation (\ref{AA0}). Now, writing $\D_{n-r}(j) = \sum_{m} \D_{n-r,m} j^m$, we can rewrite (\ref{diff-hierarchy}) as
$$
0 = \sum_{r=0}^n \sum_{j,m} a_{j,r} \D_{n-r,m} j^m y^j = \sum_{r=0}^n \sum_{j,m} a_{j,r} \D_{n-r,m} (y\partial_y)^m y^j =
\sum_{r=0}^n \Big( \sum_m \D_{n-r,m}  (y\partial_y)^m \Big) A_r  .
$$
The expression in the last bracket is nothing but the operator $\D_{n-r}(j)$ from (\ref{Sigma-def2}) with all $j$ replaced by $y\partial_y=\partial_v$. Therefore we denote this operators by $\D_{n-r}(\partial_v)$, or simply $\D_{n-r}$; for example
$$
\D_1 = \frac{S''_0}{2}  (y\partial_y)^2 + S'_1 (y\partial_y),
$$
etc. In terms of these new operators, the hierarchy of equations (\ref{diff-hierarchy}) takes a particularly simple form
\be
0 = \sum_{r=0}^n \D_{n-r} A_r,    \label{hier-SA-Cstar2}
\ee
as advertised in (\ref{hierarchy-SA}), and with $\D_{n-r}$ defined as in (\ref{Sigma-def2}) with $j$ replaced by $\partial_v$.


\subsection{Hierarchy in the $\C$ case: differential equation}

Now we show that the hierarchy of equations which arises for curves in $\C\times\C$ takes the same form (\ref{hierarchy-SA}) as in $\C^*\times\C^*$ case, even though the explicit derivation of this hierarchy is much different. Now the equation (\ref{diff-eq}) takes a form
\be
0 = \widehat{A} Z(u) = \sum_{j=0}^d \sum_{l=0}^{\infty} a_{j,l} \hbar^{l+j} \partial_u^j Z(u),  \nonumber 
\ee
and by induction we find that the last term can be written as $\partial_u^j Z = Z(\partial_u+ S')^j S'$. Then the factor of $Z$ can be factored out of an entire expression, which results in
\be
0 = \sum_{l=0}^{\infty} \Big[a_{0,l} \hbar^l + \sum_{j=0}^{d-1} a_{j+1,l} \hbar^l \Big(\hbar\partial_u + \sum_{k=0}^{\infty} \hbar^k S'_k \Big)^j \sum_{r=0}^{\infty} \hbar^r S'_r  \Big].   \label{diff-eq-Caux}
\ee
Recalling that $S'_0=v$, an explicit computation reveals that the last term in this expression can be written as
\be
\big(\hbar \partial_u + \hbar S'\big)^j \hbar S'=  v^{j+1} + \hbar\Big(S''_0 \frac{j(j+1)}{2} v^{j-1} + S'_1 (j+1)v^j  \Big) +
\ee
$$
 + \hbar^2\Big( (S''_0)^2 \frac{(j-2)(j-1)j(j+1)}{8} v^{j-3} + \big(S'''_0 + 3S''_0 S'_1\big)\frac{(j-1)j(j+1)}{6}v^{j-2} +
$$
$$
+ \big(S''_1 + (S'_1)^2 \big) \frac{j(j+1)}{2} v^{j-1} + S'_2 (j+1) v^j \Big) + \mathcal{O}(\hbar^3) =
$$
$$
= \Big[1 + \hbar\Big( \frac{S''_0}{2} \partial_v^2 + S'_1 \partial_v \Big) +
\hbar^2\Big( \frac{(S''_0)^2}{8} \partial_v^4 + \frac{S'''_0 + 3S''_0 S'_1}{6}\partial_v^3
+ \frac{S''_1 + (S'_1)^2 }{2} \partial_v^2 + S'_2 \partial_v \Big) + \mathcal{O}(\hbar^3)
  \Big]v^{j+1}.
$$
We see that a coefficient at each power $\hbar^r$ above is nothing but $\D_r$ introduced in (\ref{hier-SA-Cstar2}), i.e. the operator defined in (\ref{Sigma-def2}) with $j$ replaced by $\partial_v$.
Therefore
$$
\big(\hbar \partial_u + \hbar S'\big)^j \hbar S' = \sum_{r=0}^{\infty} \hbar^r \D_r.
$$
Using a definition $A_r$ from (\ref{Al}) we find that (\ref{diff-eq-Caux}) takes form
$$
0 = \sum_{r,l=0} \sum_{j=0}^d a_{j,l} \hbar^l  \hbar^r \D_r v^j = \sum_{r,l} \hbar^{r+l} \D_r A_l
= \sum_{n=0}^{\infty} \hbar^n \Big( \sum_{r=0}^n \D_{n-r} A_r \Big).
$$
Therefore at order $\hbar^n$ we get
\be
0 = \sum_{r=0}^n \D_{n-r} A_r,
\ee
with $\D_{n-r}$ defined as in (\ref{Sigma-def2}) with $j$ replaced by $\partial_v$. This is the same equation as in $\C^*\times\C^*$ case (\ref{hier-SA-Cstar2}), and as already advertised in (\ref{hierarchy-SA}).


\section{Quantum dilogarithm}

In literature several representations of quantum dilogarithm can be found. We use the following one
\bea
\psi(x) & = & \prod_{k=1}^{\infty} (1 - x e^{\hbar(k-1/2)})^{-1} =  \label{dilog} \\
& = & \exp \Big(-\sum_{k=1}^{\infty} \frac{x^k}{k(e^{\hbar k/2} - e^{-\hbar k/2})}   \Big) = \nonumber  \\
& = & \sum_{k=0}^{\infty} x^k e^{\frac{\hbar k}{2}} \prod_{i=1}^k \frac{1}{1 - e^{i\hbar}}, \nonumber
\eea
which has the following ``genus expansion''
\bea
\log \psi(x) & = &  \frac{1}{\hbar} S_0(x) + S_1(x) +  \hbar S_2(x) + \hbar^2 S_3(x) + \hbar^3 S_4(x) + \hbar^4 S_5(x) + \ldots   \nonumber \\
& \equiv & - \frac{1}{\hbar} \Li_2(x) + \frac{\hbar}{24} \Li_0(x) -\frac{7\hbar^3}{5760}\Li_{-2}(x) + \frac{31\hbar^5}{967680}\Li_{-4}(x) + \ldots  =  \label{qdilog-hbar}\\
& = & \sum_{k=0}^{\infty} \hbar^{k-1} (1 - 2^{1-k})\frac{B_k}{k!} \Li_{2-k}(x) \,.
\eea
Note, all terms with even power of $\hbar$ vanish. For terms $\sim \hbar^{k-1} B_k$ with $k=3,5,7,\ldots$ this is so, because $B_3=B_5=B_7=\ldots=0$. On the other hand, the term with $k=1$ is proportional to $(1-2^{1-1})=0$, hence it vanishes as well. Further details can be found {\it e.g.} in \cite{DGLZ}.


\bibliographystyle{JHEP_TD}
\bibliography{abmodel-JHEPv2}

\end{document}